\documentclass[onecolumn, peerreview]{IEEEtran}

\usepackage{bm}
\usepackage{amssymb}
\usepackage[noadjust]{cite}
\usepackage{graphicx}

\newcommand{\relvar}[2]{\buildrel {#2} \over {#1}}
\newcommand{\eqvar}[1]{\relvar{=}{#1}}

\newcommand{\levar}[1]{\relvar{\le}{#1}}
\newcommand{\gevar}[1]{\relvar{\ge}{#1}}
\newcommand{\eqdef}{\eqvar{\scriptscriptstyle\triangle}}

\newcommand{\set}[1]{\left\{ #1 \right\}}

\newcommand{\ceil}[1]{\left\lceil {#1} \right\rceil}

\newcommand{\eqref}[1]{(\textrm{\ref{#1}})}

\newcommand{\plimsup}{\mathrm{p}\mbox{-}\limsup}
\newcommand{\pliminf}{\mathrm{p}\mbox{-}\liminf}

\newcommand{\suchthat}{\mbox{s. t. }}

\newtheorem{theorem}{Theorem}
\newtheorem{corollary}{Corollary}

\newtheorem{remark}{Remark}
\newtheorem{lemma}{Lemma}

\renewcommand{\QED}{\QEDopen}
\newenvironment{proofof}[2][Proof of]{\noindent\hspace{2em}{\itshape #1 {#2}: }}{\hspace*{\fill}~\QED\par\endtrivlist\unskip}

\begin{document}

\title{An Information-Spectrum Approach to Multiterminal Rate-Distortion Theory}
\author{Shengtian Yang,
        Peiliang Qiu
\thanks{This work was supported in part by the Natural Science Foundation of China under Grant NSFC-60472079 and by the Chinese Specialized Research Fund for the Doctoral Program of Higher Education under Grant 2004-0335099.}
\thanks{S. Yang and P. Qiu are with the Department of Information Science \& Electronic Engineering, Zhejiang University, Hangzhou, 310027, China(email: yangshengtian@zju.edu.cn; qiupl@zju.edu.cn).}}

\markboth{Submitted to IEEE Transactions on Information Theory for Peer Review (April 28, 2006)}{S. Yang \MakeLowercase{\textit{et al.}}: An Information-Spectrum Approach to Multiterminal Rate-Distortion Theory}


\maketitle

\begin{abstract}
An information-spectrum approach is applied to solve the multiterminal source coding problem for correlated general sources, where sources may be nonstationary and/or nonergodic, and the distortion measure is arbitrary and may be nonadditive. A general formula for the rate-distortion region of the multiterminal source coding problem with the maximum distortion criterion under fixed-length coding is shown in this correspondence.
\end{abstract}


\IEEEpeerreviewmaketitle

\section{Introduction}

\begin{figure}[htbp]
\centering
\includegraphics[width=4in]{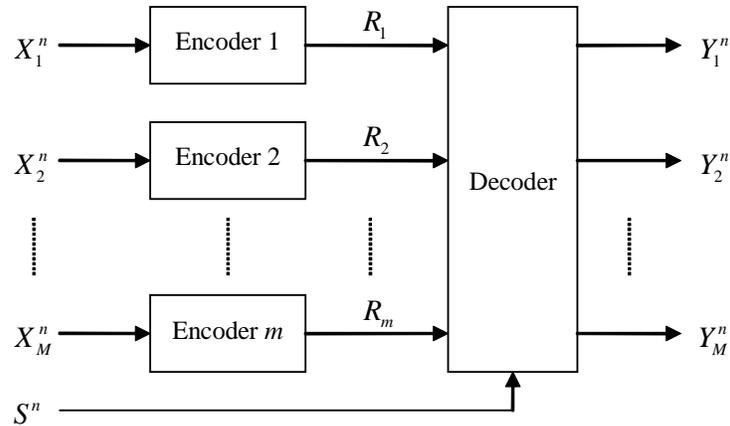}
\caption{Separate compression of $M$ correlated general sources with side information at the decoder}
\label{fig:Problem}
\end{figure}

In this correspondence, we study the classic problem in multiterminal rate-distortion theory, i.e., multiterminal source coding problem. In this problem, $M$ ($M \ge 2$) correlated general sources have to be compressed separately from each other in a lossy fashion, i.e., with respect to a fidelity criterion, and then decoded by the common decoder which has access to a side information source that is correlated with the sources to be compressed. This situation is illustrated in Fig. \ref{fig:Problem}, and it is also called distributed source coding.
The well-known Slepian-Wolf coding problem and the Wyner-Ziv coding problem can be regarded as two special cases of this situation. These two special cases were solved in 1970's for stationary memoryless sources \cite{MSC:Slepian197307, MSC:Wyner197601}, and later extended to the case of general sources \cite{MSC:Miyake199509, MSC:Iwata200206}. However, for this general problem, no conclusive results are available to date. Even for the special case that the sources are memoryless and stationary and the distortion measure is additive, only inner and outer bounds are derived in \cite{MSC:Berger197707, MSC:Gastpar200411}, etc. In this correspondence, we adopt an information-spectrum approach to solve this open problem for general sources under maximum distortion criterions. We obtain the rate-distortion region for correlated general sources, which is the main contribution of this correspondence.

The rest of this correspondence is organized as follows. In Section \ref{sec:Notations}, we first briefly introduce required notations and definitions in information-spectrum methods \cite{MSC:Han200300}, and then formally state the multiterminal source coding problem. In Section \ref{sec:MainResults}, the main theorem concerning the rate-distortion region is presented and discussed. All the proofs are finally given in Section \ref{sec:Proofs}.

\section{Notations and Definitions}\label{sec:Notations}

A general source $\bm{X}$ with alphabet $\mathcal{X}$ is characterize by an infinite sequence
$$
\{X^n = (X_1^{(n)}, X_2^{(n)}, \cdots, X_n^{(n)})\}_{n=1}^\infty
$$
of $n$-dimensional random variables $X^n$ taking values in the $n$-th Cartesian product $\mathcal{X}^n$, and in this correspondence, all the alphabets are assumed to be finite. Specifically, for $M$ ($M \ge 2$) correlated general sources, each general source $\bm{X}_m$ ($1 \le m \le M$) with alphabet $\mathcal{X}_m$ is an infinite sequence denoted by
$$
\{X_m^n = (X_{m, 1}^{(n)}, X_{m, 2}^{(n)}, \cdots, X_{m, n}^{(n)})\}_{n=1}^\infty,
$$
and the whole group of correlated general sources is denoted by $(\bm{X}_m)_{m \in \mathcal{I}_{M}}$, where $\mathcal{I}_{M}$ denotes the set $\{1, 2, \cdots, M\}$. Analogously, any part of $(\bm{X}_m)_{m \in \mathcal{I}_{M}}$ is denoted by $(\bm{X}_m)_{m \in A}$, where $A \subseteq \mathcal{I}_{M}$. In most situations, $A$ is also assumed to be an ordered set, hence $(\bm{X}_m)_{m \in A}$ is virtually a vector. Similar notations apply to any related quantities or functions, e.g., $(X_m^n)_{m \in A}$ and $(f^{(m)}(X_m^n))_{m \in A}$.

For a sequence of real random variables $\{Z_n\}_{n=1}^\infty$, the limit superior in probability $\plimsup_{n \to \infty} Z_n$ and limit inferior in probability $\pliminf_{n \to \infty} Z_n$ are defined by
$$
\plimsup_{n \to \infty} Z_n \eqdef \inf\set{\alpha \left| \lim_{n \to \infty} \Pr\{Z_n > \alpha\} = 0 \right.}
$$
and
$$
\pliminf_{n \to \infty} Z_n \eqdef \sup\set{\alpha \left| \lim_{n \to \infty} \Pr\{Z_n < \alpha\} = 0 \right.}
$$
respectively. Then for general sources $\bm{X}$, $\bm{Y}$, $\bm{\hat{X}}$ and $\bm{\hat{Y}}$, the spectral sup-entropy rate $\overline{H}(\bm{X})$, the spectral conditional sup-entropy rate $\overline{H}(\bm{X}|\bm{Y})$, the spectral sup-mutual information rate $\overline{I}(\bm{X}; \bm{Y})$, the spectral inf-mutual information rate $\underline{I}(\bm{X}; \bm{Y})$ and the spectral inf-divergence rate of $\bm{X}$ conditioned on $\bm{Y}$ with respect to $\bm{\hat{X}}$ conditioned on $\bm{\hat{Y}}$ are defined respectively by
\begin{equation}
\overline{H}(\bm{X}) \eqdef \plimsup_{n \to \infty} \frac{1}{n} \ln \frac{1}{P_{X^n}(X^n)},
\end{equation}
\begin{equation}
\overline{H}(\bm{X}|\bm{Y}) \eqdef \plimsup_{n \to \infty} \frac{1}{n} \ln \frac{1}{P_{X^n|Y^n}(X^n|Y^n)},
\end{equation}
\begin{equation}
\overline{I}(\bm{X}; \bm{Y}) \eqdef \plimsup_{n \to \infty} \frac{1}{n} \ln \frac{P_{X^nY^n}(X^n, Y^n)}{P_{X^n}(X^n) P_{Y^n}(Y^n)},
\end{equation}
\begin{equation}\label{eq:DefinitionOfInfI}
\underline{I}(\bm{X}; \bm{Y}) \eqdef \pliminf_{n \to \infty} \frac{1}{n} \ln \frac{P_{X^nY^n}(X^n, Y^n)}{P_{X^n}(X^n) P_{Y^n}(Y^n)},
\end{equation}
\begin{equation}\label{eq:DefinitionOfInfDx}
\underline{D}(\bm{X}|\bm{Y} \| \bm{\hat{X}}|\bm{\hat{Y}}) \eqdef \pliminf_{n \to \infty} \frac{1}{n} \ln \frac{P_{X^n|Y^n}(X^n|Y^n)}{P_{\hat{X}^n|\hat{Y}^n}(X^n|Y^n)}.
\end{equation}
In this correspondence, all the logarithms are calculated to the natural base $e$. Also note that in these definitions, the general source can be replaced by a sequence of random variables with arbitrary alphabets, and we denote such a sequence by $\bm{Z} = \{Z_n\}_{n=1}^\infty$ with alphabets $\bm{\mathcal{Z}} = \{\mathcal{Z}_n\}_{n=1}^\infty$, where $Z_n$ takes values in $\mathcal{Z}_n$. We also call it a general source.

If a general source is a process, that is, it satisfies the consistency condition, the notation $\{X^n\}_{n=1}^\infty$ is then replaced by the usual notation $\{X_i\}_{i=1}^\infty$ for a process and $X^n = (X_1, X_2, \cdots, X_n)$. If the processes $(\bm{X}, \bm{Y})$ are stationary and ergodic, the quantities $\overline{H}(\bm{X})$, $\overline{H}(\bm{X}|\bm{Y})$, $\overline{I}(\bm{X}; \bm{Y})$ and $\underline{I}(\bm{X}; \bm{Y})$ are then given by
$$
\overline{H}(\bm{X}) = \lim_{n \to \infty} \frac{1}{n} H(X^n),
$$
$$
\overline{H}(\bm{X}|\bm{Y}) = \lim_{n \to \infty} \frac{1}{n} H(X^n|Y^n),
$$
$$
\overline{I}(\bm{X}; \bm{Y}) = \underline{I}(\bm{X}; \bm{Y}) = \lim_{n \to \infty} \frac{1}{n} I(X^n; Y^n),
$$
where
\begin{equation}
H(X^n) \eqdef E\left[\ln \frac{1}{P_{X^n}(X^n)}\right],
\end{equation}
\begin{equation}
H(X^n|Y^n) \eqdef E\left[\ln \frac{1}{P_{X^n|Y^n}(X^n|Y^n)}\right],
\end{equation}
\begin{equation}
I(X^n;Y^n) \eqdef H(X^n) - H(X^n|Y^n).
\end{equation}
Moreover, if the stationary ergodic sources $(\bm{X}, \bm{Y})$ are memoryless, we denote them by $(X, Y)$ for convenience, and
$$
\overline{H}(X) = H(X),
$$
$$
\overline{H}(X|Y) = H(X|Y),
$$
$$
\overline{I}(X; Y) = \underline{I}(X; Y) = I(X; Y)
$$
due to the memoryless property.

The multiterminal source coding system can be stated as follows. Given correlated general sources $(\bm{X}_m)_{m \in \mathcal{I}_{M}}$ and the side information source $\bm{S}$, the $n$-length source outputs $(X_m^n)_{m \in \mathcal{I}_M}$ are separately encoded into a group of fixed-length codewords $(\phi_n^{(m)}(X_m^n))_{m \in \mathcal{I}_M}$. Then the common decoder observes these codewords $(\phi_n^{(m)}(X_m^n))_{m \in \mathcal{I}_M}$ and the side information $S^n$ to reproduce the estimates $(Y_m^n)_{m \in \mathcal{I}_M} = (\psi_n^{(m)}(S^n, (\phi_n^{(m)}(X_m^n))_{m \in \mathcal{I}_{M}}))_{m \in \mathcal{I}_M}$ of $X_m^n$. Accordingly the reproduced sequences form a group of general sources, namely, $(\bm{Y}_m)_{m \in \mathcal{I}_M}$ with reproduction alphabets $(\mathcal{Y}_m)_{m \in \mathcal{I}_M}$. Here, the encoder $(\phi_n^{(m)})_{m \in \mathcal{I}_M}$ is defined by
$$
\phi_n^{(m)}: \mathcal{X}_m^n \to \mathcal{I}_{L_n^{(m)}} \eqdef \{1, 2, \cdots, L_n^{(m)}\}
$$
and we denote the sequence $\{\phi_n^{(m)}\}_{n=1}^\infty$ and $\{\phi_n = (\phi_n^{(m)})_{m \in \mathcal{I}_M}\}_{n=1}^\infty$ by $\bm{\phi}^{(m)}$ and $\bm{\phi}$ respectively. The rate of each encoder $\phi_n^{(m)}$ is defined by
$$
R(\phi_n^{(m)}) \eqdef \frac{\ln |\phi_n^{(m)}(\mathcal{X}_m^n)|}{n},
$$
where $|A|$ denotes the cardinality of the set $A$. The decoder $(\psi_n^{(m)})_{m \in \mathcal{I}_M}$ is defined by
$$
\psi_n^{(m)}: \mathcal{S}^n \times \prod_{m' \in \mathcal{I}_M} \mathcal{I}_{L_n^{(m')}} \to \mathcal{Y}_m^n.
$$
and we denote the sequence $\{\psi_n^{(m)}\}_{n=1}^\infty$ and $\{\psi_n = (\psi_n^{(m)})_{m \in \mathcal{I}_M}\}_{n=1}^\infty$ by $\bm{\psi}^{(m)}$ and $\bm{\psi}$ respectively.

Next, let us define a general distortion measure. A general distortion measure $(\bm{d}^{(k)})_{k \in \mathcal{I}_{K}}$ is a group of sequences $\bm{d}^{(k)} = \{d_n^{(k)}\}_{n=1}^\infty$ of (measurable) functions $d_n^{(k)}$ defined by
$$
d_n^{(k)}: \prod_{m \in \mathcal{I}_{M}} \mathcal{X}_m^n \times \prod_{m \in \mathcal{I}_{M}} \mathcal{Y}_m^n \to [0, +\infty),
$$
where $K$ is a positive constant integer and $\mathcal{I}_{K} \eqdef \{1, 2, \cdots, K\}$. Then the fixed-length coding problem under the maximum distortion criterion is formulated as follows.

A rate-distortion tuple $((R_m)_{m \in \mathcal{\mathcal{I}_{M}}}, (D_k)_{k \in \mathcal{\mathcal{I}_{K}}})$ is $fm$-achievable if and only if there exists a sequence $(\bm{\phi}, \bm{\psi})$ of fixed-length codes such that
$$
(\limsup_{n \to \infty} R(\phi_n^{(m)}))_{m \in \mathcal{I}_{M}} \le (R_m)_{m \in \mathcal{I}_{M}},
$$
$$
(\plimsup_{n \to \infty} d_n^{(k)}((X_m^n)_{m \in \mathcal{I}_{M}}, \psi_n(S^n, (\phi_n^{(m)}(X_m^n))_{m \in \mathcal{I}_{M}})))_{k \in \mathcal{I}_{K}} \le (D_k)_{k \in \mathcal{I}_{K}}.
$$
Here, for any $(x_m)_{m \in \mathcal{I}_{M}}, (y_m)_{m \in \mathcal{I}_{M}} \in \mathbb{R}^M$, we say that $(x_m)_{m \in \mathcal{I}_{M}} \le (y_m)_{m \in \mathcal{I}_{M}}$ if and only if $x_m \le y_m$ for all $m \in \mathcal{I}_{M}$. Accordingly, the $fm$-rate-distortion region and $fm$-distortion-rate region are defined respectively by
$$
\mathcal{R}_{fm}((D_k)_{k \in \mathcal{I}_{K}}|(\bm{X}_m)_{m \in \mathcal{I}_M}, \bm{S}) \eqdef \{(R_m)_{m \in \mathcal{I}_{M}} | \mbox{$((R_m)_{m \in \mathcal{I}_{M}}, (D_k)_{k \in \mathcal{I}_{K}})$ is $fm$-achievable}\}
$$
and
$$
\mathcal{D}_{fm}((R_m)_{m \in \mathcal{I}_M}|(\bm{X}_m)_{m \in \mathcal{I}_M}, \bm{S}) \eqdef \{(D_k)_{k \in \mathcal{I}_{K}} | \mbox{$((R_m)_{m \in \mathcal{I}_M}, (D_k)_{k \in \mathcal{I}_K})$ is $fm$-achievable}\}.
$$
The fixed-length coding problem under the average distortion criterion can be defined analogously. A rate-distortion tuple $((R_m)_{m \in \mathcal{\mathcal{I}_{M}}}, (D_k)_{k \in \mathcal{\mathcal{I}_{K}}})$ is $fa$-achievable if and only if there exists a sequence $(\bm{\phi}, \bm{\psi})$ of fixed-length codes such that
$$
(\limsup_{n \to \infty} R(\phi_n^{(m)}))_{m \in \mathcal{I}_{M}} \le (R_m)_{m \in \mathcal{I}_{M}},
$$
$$
(\limsup_{n \to \infty} E[d_n^{(k)}((X_m^n)_{m \in \mathcal{I}_{M}}, \psi_n(S^n, (\phi_n^{(m)}(X_m^n))_{m \in \mathcal{I}_{M}}))])_{k \in \mathcal{I}_{K}} \le (D_k)_{k \in \mathcal{I}_{K}}.
$$
It can be easily shown that $fm$-achievability always implies $fa$-achievability for bounded distortion measures, and that $fa$-achievability implies $fm$-achievability for those bounded distortion measures satisfying \eqref{eq:MemorylessDistortion} defined in Section \ref{sec:MainResults} when the sources are stationary and memoryless.

\section{Main Results}\label{sec:MainResults}

In this section, we investigate the sufficient and necessary condition on the $fm$-achievability of rate-distortion tuples, thus determining the $fm$-rate-distortion region and $fm$-distortion-rate region of multiterminal source coding for correlated general sources. The main result is stated in the following theorem and the proof is presented in Section \ref{sec:Proofs}.

\begin{theorem}\label{th:AchievableCondition}
For correlated general sources $(\bm{X}_m)_{m \in \mathcal{I}_{M}}$, a side information source $\bm{S}$ and distortion measures $(\bm{d}^{(k)})_{k \in \mathcal{I}_{K}}$, the rate-distortion tuple $((R_m)_{m \in \mathcal{I}_{M}}, (D_k)_{k \in \mathcal{I}_{K}})$ is $fm$-achievable if and only if there exist general sources $(\bm{Z}^{(m)})_{m \in \mathcal{I}_{M}}$ with alphabet $\prod_{m \in \mathcal{I}_{M}} \bm{\mathcal{Z}}^{(m)} = \{\prod_{m \in \mathcal{I}_{M}} \mathcal{Z}_n^{(m)}\}_{n=1}^\infty$ and a sequence $\bm{h} = \{h_n\}_{n=1}^\infty$ of functions $h_n = (h_n^{(m)})_{m \in \mathcal{I}_{M}}$ defined by
$$
h_n^{(m)}: \mathcal{S}^n \times \prod_{m' \in \mathcal{I}_{M}} \mathcal{Z}_n^{(m')} \to \mathcal{Y}_m^n
$$
such that
\begin{equation}\label{eq:AchievableCondition1}
P_{(X_m^n)_{m \in \mathcal{I}_{M}} S^n (Z_n^{(m)})_{m \in \mathcal{I}_{M}}} = P_{(X_m^n)_{m \in \mathcal{I}_{M}} S^n} \prod_{m \in \mathcal{I}_{M}} P_{Z_n^{(m)} | X_m^n}, \quad \forall n \ge 1
\end{equation}
\begin{equation}\label{eq:AchievableCondition2}
(\plimsup_{n \to \infty} d_n^{(k)}((X_m^n)_{m \in \mathcal{I}_{M}}, h_n(S^n, (Z_n^{(m)})_{m \in \mathcal{I}_{M}})))_{k \in \mathcal{I}_{K}} \le (D_k)_{k \in \mathcal{I}_{K}},
\end{equation}
and
\begin{equation}\label{eq:AchievableCondition3}
\sum_{m \in A} R_m \ge \sum_{m \in A} \overline{I}(\bm{X}_m; \bm{Z}^{(m)}) - \underline{I}^{|A|}((\bm{Z}^{(m)})_{m \in A}) - \underline{I}((\bm{Z}^{(m)})_{m \in A}; \bm{S}, (\bm{Z}^{(m)})_{m \in \mathcal{I}_{M} \backslash A})
\end{equation}
for any nonempty set $A \subseteq \mathcal{I}_{M}$, where
\begin{equation}\label{eq:DefinitionOfInfIx}
\underline{I}^{|A|}((\bm{Z}^{(m)})_{m \in A}) \eqdef \pliminf_{n \to \infty} \frac{1}{n} \ln \frac{P_{(Z_n^{(m)})_{m \in A}}((Z_n^{(m)})_{m \in A})}{\prod_{m \in A} P_{Z_n^{(m)}}(Z_n^{(m)})}.
\end{equation}
\end{theorem}

\medskip

\begin{remark}
In Theorem \ref{th:AchievableCondition}, we introduce a new quantity $\underline{I}^{|A|}((\bm{Z}^{(m)})_{m \in A})$, which may be regarded as a generalized version of the spectral inf-mutual information rate. Also note that $\underline{I}^{|A|}((\bm{Z}^{(m)})_{m \in A}) = 0$ when $|A| = 1$.
\end{remark}

\begin{remark}
The condition \eqref{eq:AchievableCondition1} may be loosened to the following one
\begin{equation}
\alpha_n P_{(X_m^n)_{m \in \mathcal{I}_{M}} S^n} \prod_{m \in \mathcal{I}_{M}} P_{Z_n^{(m)} | X_m^n} \le P_{(X_m^n)_{m \in \mathcal{I}_{M}} S^n (Z_n^{(m)})_{m \in \mathcal{I}_{M}}} \le \beta_n P_{(X_m^n)_{m \in \mathcal{I}_{M}} S^n} \prod_{m \in \mathcal{I}_{M}} P_{Z_n^{(m)} | X_m^n}
\end{equation}
with $\lim_{n \to \infty} \alpha_n = \lim_{n \to \infty} \beta_n = 1$ (see Lemma \ref{le:UVW} in Section \ref{sec:Proofs}).
\end{remark}

By Theorem \ref{th:AchievableCondition}, the $fm$-rate-distortion region and $fm$-distortion-rate region are determined as follows.

\begin{corollary}\label{co:RateDistortionRegion}
The $fm$-rate-distortion region for a constant distortion tuple $(D_k)_{k \in \mathcal{K}}$ is
\begin{equation}
\mathcal{R}_{fm}((D_k)_{k \in \mathcal{I}_{K}}|(\bm{X}_m)_{m \in \mathcal{I}_M}, \bm{S}) = \bigcup_{(\bm{Z}^{(m)})_{m \in \mathcal{I}_{M}}} \mathcal{R}_{(\bm{Z}^{(m)})_{m \in \mathcal{I}_{M}}}((D_k)_{k \in \mathcal{I}_{K}}),
\end{equation}
where
\begin{IEEEeqnarray*}{rCl}
\mathcal{R}_{(\bm{Z}^{(m)})_{m \in \mathcal{I}_{M}}}((D_k)_{k \in \mathcal{I}_{K}}) &\eqdef &\biggl\{(R_m)_{m \in \mathcal{I}_{M}} \biggl| \sum_{m \in A} R_m \ge \sum_{m \in A} \overline{I}(\bm{X}_m; \bm{Z}^{(m)}) - \underline{I}^{|A|}((\bm{Z}^{(m)})_{m \in A}) \\
& &-\: \underline{I}((\bm{Z}^{(m)})_{m \in A}; \bm{S}, (\bm{Z}^{(m)})_{m \in \mathcal{I}_{M} \backslash A}), \mbox{ for any nonempty set $A \subseteq \mathcal{I}_{M}$} \biggr\},
\end{IEEEeqnarray*}
and $(\bm{Z}^{(m)})_{m \in \mathcal{I}_{M}}$ denotes all general sources that satisfy the conditions \eqref{eq:AchievableCondition1} and \eqref{eq:AchievableCondition2} with some sequence $\bm{h}$ of functions $h_n$.

The $fm$-distortion-rate region for a constant rate tuple $(R_m)_{m \in \mathcal{I}_M}$ is
\begin{equation}
\mathcal{D}_{fm}((R_m)_{m \in \mathcal{I}_M}|(\bm{X}_m)_{m \in \mathcal{I}_M}, \bm{S}) = \bigcup_{(\bm{Z}^{(m)})_{m \in \mathcal{I}_{M}}} \mathcal{D}_{(\bm{Z}^{(m)})_{m \in \mathcal{I}_{M}}}((R_m)_{m \in \mathcal{I}_M}),
\end{equation}
where
\begin{IEEEeqnarray*}{rCl}
\mathcal{D}_{(\bm{Z}^{(m)})_{m \in \mathcal{I}_{M}}}((R_m)_{m \in \mathcal{I}_M}) &\eqdef &\biggl\{(D_k)_{k \in \mathcal{I}_K} \biggl| (D_k)_{k \in \mathcal{I}_{K}} \ge (\plimsup_{n \to \infty} d_n^{(k)}((X_m^n)_{m \in \mathcal{I}_{M}}, h_n(S^n, (Z_n^{(m)})_{m \in \mathcal{I}_{M}})))_{k \in \mathcal{I}_{K}} \\
& &\mbox{for some sequence $\bm{h}$ of functions $h_n$} \biggr\},
\end{IEEEeqnarray*}
and $(\bm{Z}^{(m)})_{m \in \mathcal{I}_{M}}$ denotes all general sources that satisfy the conditions \eqref{eq:AchievableCondition1} and \eqref{eq:AchievableCondition3}.
\end{corollary}

To have an insight into Theorem \ref{th:AchievableCondition}, we consider some special cases. First consider the case of one terminal with side information at the decoder, we then get the rate-distortion function of the Wyner-Ziv problem for general sources.

\begin{corollary}[{\cite[Theorem 1]{MSC:Iwata200206}}]
For a general source $\bm{X}$, a side information source $S$ and distortion measures $(\bm{d}^{(k)})_{k \in \mathcal{I}_{K}}$, the rate-distortion tuple $(R, (D_k)_{k \in \mathcal{I}_{K}})$ is $fm$-achievable if and only if there exist general sources $\bm{Z}$ with alphabet $\{\mathcal{Z}_n\}_{n=1}^\infty$ and a sequence $\bm{h} = \{h_n\}_{n=1}^\infty$ of functions defined by
$$
h_n: \mathcal{S}^n \times \mathcal{Z}_n \to \mathcal{Y}^n,
$$
such that
\begin{equation}\label{eq:AchievableCondition1ForCo20}
P_{X^nS^nZ_n} = P_{X^nS^n} P_{Z_n | X^n}, \quad \forall n \ge 1
\end{equation}
\begin{equation}\label{eq:AchievableCondition2ForCo20}
(\plimsup_{n \to \infty} d_n^{(k)}(X^n, h_n(S^n, Z_n)))_{k \in \mathcal{I}_{K}} \le (D_k)_{k \in \mathcal{I}_{K}},
\end{equation}
\begin{equation}\label{eq:AchievableCondition31ForCo20}
R \ge \overline{I}(\bm{X}; \bm{Z}) - \underline{I}(\bm{Z}; \bm{S}).
\end{equation}
and hence the infimum of the achievable rate for a constant distortion tuple $(D_{k})_{k \in \mathcal{I}_{K}}$ is given by
\begin{equation}
\inf \{\overline{I}(\bm{X}; \bm{Z}) - \underline{I}(\bm{Z}; \bm{S})\},
\end{equation}
where $\inf$ is over all $\bm{Z}$ and $\{h_n\}_{n=1}^\infty$ satisfying properties \eqref{eq:AchievableCondition1ForCo20} and \eqref{eq:AchievableCondition2ForCo20}.
\end{corollary}

Second, let us consider the case of two terminals without side information at the decoder. Then Theorem \ref{th:AchievableCondition} is reduced to the following form.

\begin{corollary}\label{co:AchievableConditionForTwoTerminals}
For correlated general sources $(\bm{X}_1, \bm{X}_2)$ and distortion measures $(\bm{d}^{(k)})_{k \in \mathcal{I}_{K}}$, the rate-distortion tuple $(R_1, R_2, \linebreak (D_k)_{k \in \mathcal{I}_{K}})$ is $fm$-achievable if and only if there exist general sources $(\bm{Z}^{(1)}, \bm{Z}^{(2)})$ with alphabet $\{\mathcal{Z}_n^{(1)} \times \mathcal{Z}_n^{(2)}\}_{n=1}^\infty$ and a sequence $\bm{h} = \{h_n\}_{n=1}^\infty$ of functions $h_n = (h_n^{(1)}, h_n^{(2)})$ defined by
$$
h_n: \mathcal{Z}_n^{(1)} \times \mathcal{Z}_n^{(2)} \to \mathcal{Y}_1^n \times \mathcal{Y}_2^n
$$
such that
\begin{equation}\label{eq:AchievableCondition1ForCo2}
P_{X_1^nX_2^nZ_n^{(1)}Z_n^{(2)}} = P_{X_1^nX_2^n} P_{Z_n^{(1)} | X_1^n} P_{Z_n^{(2)} | X_2^n},
\end{equation}
\begin{equation}\label{eq:AchievableCondition2ForCo2}
(\plimsup_{n \to \infty} d_n^{(k)}(X_1^n, X_2^n, h_n(Z_n^{(1)}, Z_n^{(2)})))_{k \in \mathcal{I}_{K}} \le (D_k)_{k \in \mathcal{I}_{K}},
\end{equation}
and
\begin{equation}\label{eq:AchievableCondition31ForCo2}
R_1 \ge \overline{I}(\bm{X}_1; \bm{Z}^{(1)}) - \underline{I}(\bm{Z}^{(1)}; \bm{Z}^{(2)}),
\end{equation}
\begin{equation}\label{eq:AchievableCondition32ForCo2}
R_2 \ge \overline{I}(\bm{X}_2; \bm{Z}^{(2)}) - \underline{I}(\bm{Z}^{(1)}; \bm{Z}^{(2)}),
\end{equation}
\begin{equation}\label{eq:AchievableCondition33ForCo2}
R_1 + R_2 \ge \overline{I}(\bm{X}_1; \bm{Z}^{(1)}) + \overline{I}(\bm{X}_2; \bm{Z}^{(2)}) - \underline{I}(\bm{Z}^{(1)}; \bm{Z}^{(2)}).
\end{equation}
\end{corollary}

\medskip

If for each $k \in \mathcal{I}_K$, the distortion measure $d_n^{(k)}$ is additive, that is, it is defined by
\begin{equation}\label{eq:MemorylessDistortion}
d_n^{(k)}(\bm{x}_1, \bm{x}_2, \bm{y}_1, \bm{y}_2) = \frac{1}{n} \sum_{i=1}^n d^{(k)}(x_{1,i}, x_{2,i}, y_{1,i}, y_{2,i}),
\end{equation}
for all $\bm{x}_1 \in \mathcal{X}_1^n$, $\bm{x}_2 \in \mathcal{X}_2^n$, $\bm{y}_1 \in \mathcal{Y}_1^n$ and $\bm{y}_2 \in \mathcal{Y}_2^n$, where $d^{(k)}$ is a nonnegative (measurable) function called memoryless distortion measure, then for stationary memoryless sources $(X_1, X_2)$, the sufficient condition given by \cite{MSC:Berger197707} follows from Corollary \ref{co:AchievableConditionForTwoTerminals}.

\begin{corollary}[{\cite[Theorem 6.1]{MSC:Berger197707}}]\label{co:OldSufficientCondition}
For correlated stationary memoryless sources $(X_1, X_2)$ and memoryless distortion measure $(d^{(k)})_{k \in \mathcal{I}_{K}}$, the rate-distortion tuple $(R_1, R_2, (D_k)_{k \in \mathcal{I}_{K}})$ is $fm$-achievable (or $fa$-achievable) if there exist a pair of random variables $(Z_1, Z_2)$ with alphabet $\mathcal{Z}_1 \times \mathcal{Z}_2$ and a pair of functions $h^{(1)}: \mathcal{Z}_1 \times \mathcal{Z}_2 \to \mathcal{Y}_1$ and $h^{(2)}: \mathcal{Z}_1 \times \mathcal{Z}_2 \to \mathcal{Y}_2$ such that
\begin{equation}\label{eq:AchievableCondition1ForCo3}
P_{X_1X_2Z_1Z_2} = P_{X_1X_2} P_{Z_1|X_1} P_{Z_2|X_2},
\end{equation}
\begin{equation}\label{eq:AchievableCondition2ForCo3}
(E[d^{(k)}(X_1, X_2, h^{(1)}(Z_1, Z_2), h^{(2)}(Z_1, Z_2))])_{k \in \mathcal{I}_{K}} \le (D_k)_{k \in \mathcal{I}_{K}}
\end{equation}
\begin{equation}\label{eq:AchievableCondition31ForCo3}
R_1 \ge I(X_1; Z_1 | Z_2), \quad R_2 \ge I(X_2; Z_2 | Z_1),
\end{equation}
\begin{equation}\label{eq:AchievableCondition32ForCo3}
R_1 + R_2 \ge I(X_1, X_2; Z_1, Z_2).
\end{equation}
\end{corollary}

\medskip

The proof is easy and hence omitted here. Note that the memoryless distortion measures are bounded since the alphabets are finite, and the sources $(X_1, X_2, Z_1, Z_2)$ are jointly stationary and memoryless, which implies the validity of the law of large number or the ergodic theorem.

Though we have determined the whole $fm$-rate-distortion or $fm$-distortion-rate region for correlated general sources under the maximum distortion criterion, this does not mean that the multiterminal source coding problem has been solved. First, the $fm$-rate-distortion or $fm$-distortion-rate region in Corollary \ref{co:RateDistortionRegion} are obviously incomputable in general. Second, even for the correlated memoryless sources, the single letter $fm$-rate-distortion (or $fa$-rate-distortion) region is still unknown. It seems that the usual treatment for memoryless sources or channels in information-spectrum methods can not be easily applied to this problem, and one of the difficulties is how to reduce the general function $h_n = (h_n^{(m)})_{m \in \mathcal{I}_{M}}$ to a memoryless form if we do not use the method in \cite{MSC:Berger197707}. But in any case, Theorem \ref{th:AchievableCondition} does provide a very general sufficient condition. and it also gives some hints on the characterization of single letter rate-distortion region. Under the same settings of Corollary \ref{co:OldSufficientCondition}, \emph{we believe that for any $fm$-achievable rate-distortion tuple, there exits a pair of random variables $(Z_1, Z_2)$ which can be regarded as a mixture of infinite general sources satisfying the condition \eqref{eq:AchievableCondition1}, and most of the general sources (with probability $1$) achieve the same rate-distortion tuple.}

Finally, let us turn our attention to the multiterminal source coding problem for the mixed sources as an application example of Theorem \ref{th:AchievableCondition}. For simplicity, we will consider the settings of Corollary \ref{co:AchievableConditionForTwoTerminals}, and we assume that the source pair is a mixed source pair.

Let $(\bm{X}_{1\alpha}, \bm{X}_{2\alpha})$ and $(\bm{X}_{1\beta}, \bm{X}_{2\beta})$ be two correlated general sources, and let us define a mixed source pair $(\bm{X}_1, \bm{X}_2)$ by
$$
P_{X_1^nX_2^n}(\bm{x}_1, \bm{x}_2) = \alpha P_{X_{1\alpha}^nX_{2\alpha}^n}(\bm{x}_1, \bm{x}_2) + \beta P_{X_{1\beta}^nX_{2\beta}^n}(\bm{x}_1, \bm{x}_2)
$$
for $\bm{x}_1 \in \mathcal{X}_1^n$, $\bm{x}_2 \in \mathcal{X}_2^n$, where $\alpha > 0$, $\beta > 0$ are constants such that $\alpha + \beta = 1$. Then for this mixed source pair, we have the following corollary derived from Corollary \ref{co:AchievableConditionForTwoTerminals}.

\begin{corollary}
For given correlated mixed sources $(\bm{X}_1, \bm{X}_2)$ with component sources $(\bm{X}_{1\alpha}, \bm{X}_{2\alpha})$ and $(\bm{X}_{1\beta}, \bm{X}_{2\beta})$ and distortion measures $(\bm{d}^{(k)})_{k \in \mathcal{I}_{K}}$, the rate-distortion tuple $(R_1, R_2, (D_k)_{k \in \mathcal{I}_{K}})$ is $fm$-achievable if and only if there exist general sources $(\bm{Z}^{(1)}, \bm{Z}^{(2)})$ with alphabet $\{\mathcal{Z}_n^{(1)} \times \mathcal{Z}_n^{(2)}\}_{n=1}^\infty$ and a sequence $\bm{h} = \{h_n\}_{n=1}^\infty$ of functions $h_n = (h_n^{(1)}, h_n^{(2)})$ defined by
$$
h_n: \mathcal{Z}_n^{(1)} \times \mathcal{Z}_n^{(2)} \to \mathcal{Y}_1^n \times \mathcal{Y}_2^n
$$
such that
\begin{equation}\label{eq:AchievableCondition1ForCo5}
P_{X_1^nX_2^nZ_n^{(1)}Z_n^{(2)}} = P_{X_1^nX_2^n} P_{Z_n^{(1)} | X_1^n} P_{Z_n^{(2)} | X_2^n},
\end{equation}
\begin{equation}\label{eq:AchievableCondition2ForCo5}
(\max\{\plimsup_{n \to \infty} d_n^{(k)}(X_{1\alpha}^n, X_{2\alpha}^n, h_n(Z_n^{(1)}, Z_n^{(2)})), \plimsup_{n \to \infty} d_n^{(k)}(X_{1\beta}^n, X_{2\beta}^n, h_n(Z_n^{(1)}, Z_n^{(2)}))\})_{k \in \mathcal{I}_{K}} \le (D_k)_{k \in \mathcal{I}_{K}},
\end{equation}
and
\begin{equation}\label{eq:AchievableCondition31ForCo5}
R_1 \ge \max\{\overline{I}(\bm{X}_{1\alpha}; \bm{Z}^{(1)}), \overline{I}(\bm{X}_{1\beta}; \bm{Z}^{(1)})\} - \underline{I}(\bm{Z}^{(1)}; \bm{Z}^{(2)}),
\end{equation}
\begin{equation}\label{eq:AchievableCondition32ForCo5}
R_2 \ge \max\{\overline{I}(\bm{X}_{2\alpha}; \bm{Z}^{(2)}), \overline{I}(\bm{X}_{2\beta}; \bm{Z}^{(2)})\} - \underline{I}(\bm{Z}^{(1)}; \bm{Z}^{(2)}),
\end{equation}
\begin{equation}\label{eq:AchievableCondition33ForCo5}
R_1 + R_2 \ge \max\{\overline{I}(\bm{X}_{1\alpha}; \bm{Z}^{(1)}), \overline{I}(\bm{X}_{1\beta}; \bm{Z}^{(1)})\} + \max\{\overline{I}(\bm{X}_{2\alpha}; \bm{Z}^{(2)}), \overline{I}(\bm{X}_{2\beta}; \bm{Z}^{(2)})\} - \underline{I}(\bm{Z}^{(1)}; \bm{Z}^{(2)}).
\end{equation}
\end{corollary}

The proof is similar to the proof of Corollary 1 in \cite{MSC:Iwata200206} and hence omitted here. Lemma 3 and 4 in \cite{MSC:Iwata200206} are applied repeatedly in the proof.

\section{Proofs of Theorems}\label{sec:Proofs}

To prove Theorem \ref{th:AchievableCondition}, we first need to establish a series of lemmas.

\begin{lemma}\label{le:InfDx}
Let $\bm{X} = \{X_n\}_{n=1}^\infty$, $\bm{\hat{X}} = \{\hat{X}_n\}_{n=1}^\infty$ and $\bm{Y} = \{Y_n\}_{n=1}^\infty$, $\bm{\hat{Y}} = \{\hat{Y}_n\}_{n=1}^\infty$ be arbitrary general sources with alphabets $\bm{\mathcal{X}} = \{\mathcal{X}_n\}_{n=1}^\infty$ and $\bm{\mathcal{Y}} = \{\mathcal{Y}_n\}_{n=1}^\infty$ respectively, then we have
\begin{equation}\label{eq:NonnegativeInfDx}
\underline{D}(\bm{X}|\bm{Y} \| \bm{\hat{X}}|\bm{\hat{Y}}) \ge 0.
\end{equation}
\end{lemma}

\begin{proof}
For any $\gamma > 0$,
\begin{IEEEeqnarray*}{rCl}
\Pr\biggl\{\frac{1}{n} \ln \frac{P_{X_n|Y_n}(X_n|Y_n)}{P_{\hat{X}_n|\hat{Y}_n}(X_n|Y_n)} < -\gamma\biggr\} &= &\sum_{\bm{x} \in \mathcal{X}_n, \bm{y} \in \mathcal{Y}_n} P_{X_nY_n}(\bm{x}, \bm{y}) 1\biggl\{\frac{P_{X_n|Y_n}(\bm{x}|\bm{y})}{P_{\hat{X}_n|\hat{Y}_n}(\bm{x}|\bm{y})} < e^{-n\gamma}\biggr\} \\
&< &\sum_{\bm{x} \in \mathcal{X}_n, \bm{y} \in \mathcal{Y}_n} e^{-n\gamma} P_{Y_n}(\bm{y}) P_{\hat{X}_n|\hat{Y}_n}(\bm{x}|\bm{y}) \\
&= &e^{-n\gamma} \sum_{\bm{y} \in \mathcal{Y}_n} P_{Y_n}(\bm{y}) \sum_{\bm{x} \in \mathcal{X}_n} P_{\hat{X}_n|\hat{Y}_n}(\bm{x}|\bm{y}) \\
&= &e^{-n\gamma},
\end{IEEEeqnarray*}
hence
$$
\lim_{n \to \infty} \Pr\{\frac{1}{n} \ln \frac{P_{X_n|Y_n}(X_n|Y_n)}{P_{\hat{X}_n|\hat{Y}_n}(X_n|Y_n)} < -\gamma\} = 0
$$
for any $\gamma > 0$, and this concludes \eqref{eq:NonnegativeInfDx} by the definition \eqref{eq:DefinitionOfInfDx}.
\end{proof}

\begin{lemma}[An enhanced version of Lemma 1 in \cite{MSC:Iwata200206}]\label{le:UVW}
Let $\bm{U} = \{U_n\}_{n=1}^\infty$, $\bm{V} = \{V_n\}_{n=1}^\infty$ and $\bm{W} = \{W_n\}_{n=1}^\infty$ be arbitrary general sources with alphabets $\bm{\mathcal{U}} = \{\mathcal{U}_n\}$, $\bm{\mathcal{V}} = \{\mathcal{V}_n\}$ and $\bm{\mathcal{W}} = \{\mathcal{W}_n\}$ respectively, and they satisfy
\begin{equation}\label{eq:MarkovConditionForLemmaUVW}
P_{U_nV_nW_n} \ge c_n P_{U_nV_n} P_{W_n|V_n}
\end{equation}
with $\lim_{n \to \infty} c_n = 1$. Now let $\{B_n\}_{n=1}^\infty$ be a sequence of arbitrary (measurable) sets in $\mathcal{U}_n \times \mathcal{W}_n$ such that
\begin{equation}\label{eq:ProbabilityOfBnForLemmaUVW}
\liminf_{n \to \infty} \Pr\{(U_n, W_n) \in B_n\} = \epsilon.
\end{equation}
Then, for any $\gamma > 0$, there exits a sequence $\bm{f} = \{f_n\}_{n=1}^\infty$ of functions $f_n: \mathcal{V}_n \to \mathcal{W}_n$ such that
$$
|f_n(\mathcal{V}_n)| \le \ceil{e^{n(\overline{I}(\bm{V}; \bm{W}) + \gamma)}},
$$
$$
\liminf_{n \to \infty} \Pr\{(U_n, f_n(V_n)) \in B_n\} \ge \epsilon.
$$
\end{lemma}

The proof of Theorem \ref{th:AchievableCondition} relies heavily on Lemma 1 in \cite{MSC:Iwata200206}, but here we choose to present an enhanced version of Lemma 1 in \cite{MSC:Iwata200206}, and its proof is greatly simplified by applying Theorem 5.5.1 in \cite{MSC:Han200300}.

\begin{proof}
Let us define a distortion measure
$$
d_n(\bm{v}, \bm{w}) = \sum_{\bm{u} \in \mathcal{U}_n} P_{U_n|V_nW_n}(\bm{u}|\bm{v}, \bm{w}) 1\{(\bm{u}, \bm{w}) \not \in B_n\}
$$
for all $\bm{v} \in \mathcal{V}_n$, $\bm{w} \in \mathcal{W}_n$. Clearly, $d_n$ is bounded, and we have
$$
\limsup_{n \to \infty} E[d_n(V_n, W_n)] = \limsup_{n \to \infty} \sum_{\bm{v} \in \mathcal{V}_n, \bm{w} \in \mathcal{W}_n} P_{V_nW_n}(\bm{v}, \bm{w}) d_n(\bm{v}, \bm{w}) = \limsup_{n \to \infty} \Pr\{(U_n, W_n) \not \in B_n\} = 1 - \epsilon
$$
by the condition \eqref{eq:ProbabilityOfBnForLemmaUVW}. Then from the proof of the direct part of Theorem 5.5.1 in \cite{MSC:Han200300}, it follows that for any $\gamma > 0$, there exists a sequence $\bm{f} = \{f_n\}_{n=1}^\infty$ of functions $f_n: \mathcal{V}_n \to \mathcal{W}_n$ such that
$$
|f_n(\mathcal{V}_n)| \le \ceil{e^{n(\overline{I}(\bm{V}; \bm{W}) + \gamma)}},
$$
$$
\limsup_{n \to \infty} E[d_n(V_n, f(V_n))] \le 1 - \epsilon.
$$
Finally, we have
\begin{IEEEeqnarray*}{rCl}
\Pr\{(U_n, f_n(V_n)) \in B_n\} &= &1 - \Pr\{(U_n, f_n(V_n)) \not \in B_n\} \\
&= &1 - \sum_{\bm{v} \in \mathcal{V}_n} P_{V_n}(\bm{v}) \sum_{\bm{u} \in \mathcal{U}_n} P_{U_n|V_n}(\bm{u}|\bm{v}) 1\{(\bm{u}, f_n(\bm{v})) \not \in B_n\} \\
&\gevar{(a)} &1 - c_n^{-1} \sum_{\bm{v} \in \mathcal{V}_n} P_{V_n}(\bm{v}) \sum_{\bm{u} \in \mathcal{U}_n} P_{U_n|V_nW_n}(\bm{u}|\bm{v}, f_n(\bm{v})) 1\{(\bm{u}, f_n(\bm{v})) \not \in B_n\} \\
&= &1 - c_n^{-1} E[d_n(V_n, f(V_n))],
\end{IEEEeqnarray*}
where (a) follows from \eqref{eq:MarkovConditionForLemmaUVW}, and hence
$$
\liminf_{n \to \infty} \Pr\{(U_n, f_n(V_n)) \in B_n\} \ge 1 - \limsup_{n \to \infty} \frac{E[d_n(V_n, f(V_n))]}{c_n} \ge 1 - \frac{\limsup_{n \to \infty} E[d_n(V_n, f(V_n))]}{\lim_{n \to \infty} c_n} \ge \epsilon.
$$
The proof is completed.
\end{proof}

\begin{lemma}\label{le:XSZ}
Let $(\bm{X}_m)_{m \in \mathcal{I}_M}$, $\bm{S}$ and $(\bm{Z}^{(m)})_{m \in \mathcal{I}_M}$ be arbitrary general sources satisfying \eqref{eq:AchievableCondition1}. Now let $\{B_n\}_{n=1}^\infty$ be a sequence of arbitrary (measurable) sets $B_n$ in $\prod_{m \in \mathcal{I}_M} \mathcal{X}_m^n \times \mathcal{S}^n \times \prod_{m \in \mathcal{I}_M} \mathcal{Z}_n^{(m)}$ such that
$$
\lim_{n \to \infty} \Pr\{((X_m^n)_{m \in \mathcal{I}_M}, S^n, (Z_n^{(m)})_{m \in \mathcal{I}_M}) \in B_n\} = 1.
$$
Then, for any $\gamma_1, \gamma_2 > 0$, there exist a group of sequences $(\bm{f}^{(m)} = \{f_n^{(m)}\}_{n=1}^\infty)_{m \in \mathcal{I}_M}$ of functions $f_n^{(m)}: \mathcal{X}_m^n \to \mathcal{Z}_n^{(m)}$ such that
$$
|f_n^{(m)}(\mathcal{X}_m^n)| \le e^{\ceil{n(\overline{I}(\bm{X}_m; \bm{Z}^{(m)}) + \gamma_1)}},
$$
$$
\lim_{n \to \infty} \Pr\{((X_m^n)_{m \in \mathcal{I}_M}, S^n, (f_n^{(m)}(X_m^n))_{m \in \mathcal{I}_M}) \in B_n\} = 1,
$$
and
$$
\underline{I}^{|A|}((\bm{f}^{(m)}(\bm{X}_m))_{m \in A}) \ge \underline{I}^{|A|}((\bm{Z}^{(m)})_{m \in A}) - \gamma_2,
$$
$$
\underline{I}((\bm{f}^{(m)}(\bm{X}_m))_{m \in A}; \bm{S}, (\bm{f}^{(m)}(\bm{X}_m))_{m \in \mathcal{I}_{M} \backslash A}) \ge \underline{I}((\bm{Z}^{(m)})_{m \in A}; \bm{S}, (\bm{Z}^{(m)})_{m \in \mathcal{I}_{M} \backslash A}) - \gamma_2
$$
for any nonempty set $A \subseteq \mathcal{I}_{M}$, where $\bm{f}^{(m)}(\bm{X}_m) \eqdef \{f_n^{(m)}(X_m^n)\}_{n=1}^\infty$.
\end{lemma}

\begin{proof}
Supposing that for a given $J$ ($0 \le J < M$), we have
\begin{equation}\label{eq:fnForLemmaXSZ}
|f_n^{(m)}(\mathcal{X}_m^n)| \le e^{\ceil{n(\overline{I}(\bm{X}_m; \bm{Z}^{(m)}) + \gamma_1)}}, \quad \forall m \in \mathcal{I}_{J}
\end{equation}
\begin{equation}\label{eq:BnInProbabilityForLemmaXSZ}
\lim_{n \to \infty} \Pr\{((X_m^n)_{m \in \mathcal{I}_M}, S^n, (f_n^{(m)}(X_m^n))_{m \in \mathcal{I}_{J}}, (Z_n^{(m)})_{m \in \mathcal{I}_{M} \backslash \mathcal{I}_{J}}) \in B_n\} = 1,
\end{equation}
and
\begin{equation}\label{eq:InfIxForJForLemmaXSZ}
\underline{I}^{|A|}((\bm{f}^{(m)}(\bm{X}_m))_{m \in A \cap \mathcal{I}_J}, (\bm{Z}^{(m)})_{m \in A \backslash \mathcal{I}_J}) \ge \underline{I}^{|A|}((\bm{Z}^{(m)})_{m \in A}) - \frac{J}{M}\gamma_2,
\end{equation}
\begin{IEEEeqnarray}{Cl}
&\underline{I}((\bm{f}^{(m)}(\bm{X}_m))_{m \in A \cap \mathcal{I}_J}, (\bm{Z}^{(m)})_{m \in A \backslash \mathcal{I}_J}; \bm{S}, (\bm{f}^{(m)}(\bm{X}_m))_{m \in (\mathcal{I}_{M} \backslash A) \cap \mathcal{I}_J}, (\bm{Z}^{(m)})_{m \in (\mathcal{I}_{M} \backslash A) \backslash \mathcal{I}_J}) \IEEEnonumber \\
\ge &\underline{I}((\bm{Z}^{(m)})_{m \in A}; \bm{S}, (\bm{Z}^{(m)})_{m \in \mathcal{I}_{M} \backslash A}) - \frac{J}{M}\gamma_2 \label{eq:InfIForLemmaXSZ}
\end{IEEEeqnarray}
for any nonempty set $A \subseteq \mathcal{I}_{M}$. Clearly, the above conditions hold trivially when $J = 0$, and the lemma corresponds to the case $J = M$. So to prove the lemma, we only need to show that the above conditions also hold for the case of $J+1$.

Now, let $U_n = ((X_m^n)_{m \in \mathcal{I}_M}, S^n, (Z_n^{(m)})_{m \in \mathcal{I}_M \backslash \mathcal{I}_{J+1}})$, $V_n = X_{J+1}^n$, $W_n = Z_n^{(J+1)}$. From \eqref{eq:AchievableCondition1}, it follows that
$$
P_{(X_m^n)_{m \in \mathcal{I}_M}S^n(Z_n^{(m)})_{m \in \mathcal{I}_M \backslash \mathcal{I}_{J}}} = P_{(X_m^n)_{m \in \mathcal{I}_M}S^n(Z_n^{(m)})_{m \in \mathcal{I}_M \backslash \mathcal{I}_{J+1}}} P_{Z_n^{(J+1)}|X_{J+1}^n},
$$
hence we have
$$
P_{U_nV_nW_n} = P_{U_nV_n}P_{W_n|V_n}.
$$
Next, we define
\begin{IEEEeqnarray*}{rCl}
C_n &= &\biggl\{((\bm{x}_m)_{m \in \mathcal{I}_M}, \bm{s}, (\bm{z}^{(m)})_{m \in \mathcal{I}_M \backslash \mathcal{I}_J}) \in \prod_{m \in \mathcal{I}_M} \mathcal{X}_m^n \times \mathcal{S}^n \times \prod_{m \in \mathcal{I}_M \backslash \mathcal{I}_J} \mathcal{Z}_n^{(m)} \bigg| \\
& &((\bm{x}_m)_{m \in \mathcal{I}_M}, \bm{s}, (f_n^{(m)}(\bm{x}_m))_{m \in \mathcal{I}_J}, (\bm{z}^{(m)})_{m \in \mathcal{I}_M \backslash \mathcal{I}_J}) \in B_n \biggr\}
\end{IEEEeqnarray*}
$$
C_n' = C_n \cap T_n^{(1)} \cap T_n^{(2)},
$$
$$
C_n'' = \{(\bm{u}, \bm{w}) \in \mathcal{U}_n \times \mathcal{W}_n | \bm{u} = ((\bm{x}_m)_{m \in \mathcal{I}_M}, \bm{s}, (\bm{z}^{(m)})_{m \in \mathcal{I}_M \backslash \mathcal{I}_{J+1}}), \bm{w} = \bm{z}^{(J+1)}, ((\bm{x}_m)_{m \in \mathcal{I}_M}, \bm{s}, (\bm{z}^{(m)})_{m \in \mathcal{I}_M \backslash \mathcal{I}_{J}}) \in C_n'\}
$$
where
\begin{IEEEeqnarray}{rCl}
T_n^{(1)} &= &\biggl\{ ((\bm{x}_m)_{m \in \mathcal{I}_M}, \bm{s}, (\bm{z}^{(m)})_{m \in \mathcal{I}_M \backslash \mathcal{I}_J}) \in \prod_{m \in \mathcal{I}_M} \mathcal{X}_m^n \times \mathcal{S}^n \times \prod_{m \in \mathcal{I}_M \backslash \mathcal{I}_J} \mathcal{Z}_n^{(m)} \bigg| \IEEEnonumber \\
& &\frac{1}{n} \ln \frac{P_{(f_n^{(m)}(X_m^n))_{m \in A \cap \mathcal{I}_J} (Z_n^{(m)})_{m \in A \backslash \mathcal{I}_J}}((f_n^{(m)}(\bm{x}_m))_{m \in A \cap \mathcal{I}_J}, (\bm{z}^{(m)})_{m \in A \backslash \mathcal{I}_J})}{\prod_{m \in A \cap \mathcal{I}_J} P_{f_n^{(m)}(X_m^n)}(f_n^{(m)}(\bm{x}_m)) \prod_{m \in A \backslash \mathcal{I}_J} P_{Z_n^{(m)}}(\bm{z}^{(m)})} \IEEEnonumber \\
& &\ge \underline{I}^{|A|}((\bm{f}^{(m)}(\bm{X}_m))_{m \in A \cap \mathcal{I}_J}, (\bm{Z}^{(m)})_{m \in A \backslash \mathcal{I}_J}) - \frac{1}{M} \gamma_2, \mbox{ for all $A$ containing $J+1$} \biggr\},\label{eq:DefinitionOfTn1ForLemmaXSZ}
\end{IEEEeqnarray}
\begin{IEEEeqnarray}{rCl}
T_n^{(2)} &= &\biggl\{ ((\bm{x}_m)_{m \in \mathcal{I}_M}, \bm{s}, (\bm{z}^{(m)})_{m \in \mathcal{I}_M \backslash \mathcal{I}_J}) \in \prod_{m \in \mathcal{I}_M} \mathcal{X}_m^n \times \mathcal{S}^n \times \prod_{m \in \mathcal{I}_M \backslash \mathcal{I}_J} \mathcal{Z}_n^{(m)} \bigg| \IEEEnonumber \\
& &\frac{1}{n} \ln \biggl[ P_{S^n (f_n^{(m)}(X_m^n))_{m \in \mathcal{I}_J} (Z_n^{(m)})_{m \in \mathcal{I}_M \backslash \mathcal{I}_J}}(\bm{s}, (f_n^{(m)}(\bm{x}_m))_{m \in \mathcal{I}_J}, (\bm{z}^{(m)})_{m \in \mathcal{I}_M \backslash \mathcal{I}_J}) \bigg/ \IEEEnonumber \\
& &\biggl( P_{(f_n^{(m)}(X_m^n))_{m \in A \cap \mathcal{I}_J} (Z_n^{(m)})_{m \in A \backslash \mathcal{I}_J}}((f_n^{(m)}(\bm{x}_m))_{m \in A \cap \mathcal{I}_J}, (\bm{z}^{(m)})_{m \in A \backslash \mathcal{I}_J}) \IEEEnonumber \\
& &P_{S^n (f_n^{(m)}(X_m^n))_{m \in (\mathcal{I}_M \backslash A) \cap \mathcal{I}_J} (Z_n^{(m)})_{m \in (\mathcal{I}_M \backslash A) \backslash \mathcal{I}_J}}(\bm{s}, (f_n^{(m)}(\bm{x}_m))_{m \in (\mathcal{I}_M \backslash A) \cap \mathcal{I}_J}, (\bm{z}^{(m)})_{m \in (\mathcal{I}_M \backslash A) \backslash \mathcal{I}_J}) \biggr) \biggr] \IEEEnonumber \\
& &\ge \underline{I}((\bm{f}^{(m)}(\bm{X}_m))_{m \in A \cap \mathcal{I}_J}, (\bm{Z}^{(m)})_{m \in A \backslash \mathcal{I}_J}; \bm{S}, (\bm{f}^{(m)}(\bm{X}_m))_{m \in (\mathcal{I}_{M} \backslash A) \cap \mathcal{I}_J}, (\bm{Z}^{(m)})_{m \in (\mathcal{I}_{M} \backslash A) \backslash \mathcal{I}_J}) - \frac{1}{M} \gamma_2, \IEEEnonumber \\
& &\mbox{for all nonempty $A \subseteq \mathcal{I}_M$} \biggr\}.\label{eq:DefinitionOfTn2ForLemmaXSZ}
\end{IEEEeqnarray}
Clearly, by the definition \eqref{eq:DefinitionOfInfI} and \eqref{eq:DefinitionOfInfIx}, we have
\begin{IEEEeqnarray*}{rCl}
\lim_{n \to \infty} \Pr\{(U_n, W_n) \in C_n''\} &= &\lim_{n \to \infty} \Pr\{((X_m^n)_{m \in \mathcal{I}_M}, S^n, (Z_n^{(m)})_{m \in \mathcal{I}_M \backslash \mathcal{I}_J}) \in C_n'\} \\
&\ge &1 - \lim_{n \to \infty} \Pr\{((X_m^n)_{m \in \mathcal{I}_M}, S^n, (f_n^{(m)}(X_m^n))_{m \in \mathcal{I}_J}, (Z_n^{(m)})_{m \in \mathcal{I}_M \backslash \mathcal{I}_J}) \not \in B_n\} \\
& &- \lim_{n \to \infty} \Pr\{((X_m^n)_{m \in \mathcal{I}_M}, S^n, (Z_n^{(m)})_{m \in \mathcal{I}_M \backslash \mathcal{I}_J}) \not \in T_n^{(1)} \cap T_n^{(2)}\} \\
&= &1,
\end{IEEEeqnarray*}
and hence $\lim_{n \to \infty} \Pr\{(U_n, W_n) \in C_n''\} = 1$. Then from Lemma \ref{le:UVW}, it follows that there exists a sequence $\bm{f}^{(J+1)} = \{f_n^{(J+1)}\}_{n=1}^\infty$ of functions $f_n^{(J+1)}: \mathcal{X}_{J+1}^n \to \mathcal{Z}_n^{(J+1)}$ such that
\begin{equation}
|f_n^{(J+1)}(\mathcal{X}_{J+1}^n)| \le \ceil{e^{n(\overline{I}(\bm{X}_{J+1}; \bm{Z}^{(J+1)}) + \gamma_1)}},
\end{equation}
\begin{equation}\label{eq:C_n'inProbabilityForLemmaXSZ}
\lim_{n \to \infty} \Pr\{((X_m^n)_{m \in \mathcal{I}_M}, S^n, f_n^{(J+1)}(X_m^n), (Z_n^{(m)})_{m \in \mathcal{I}_M \backslash \mathcal{I}_{J+1}}) \in C_n'\} = 1.
\end{equation}
Hence, we have
\begin{IEEEeqnarray*}{Cl}
&\lim_{n \to \infty} \Pr\{((X_m^n)_{m \in \mathcal{I}_M}, S^n, (f_n^{(m)}(X_m^n))_{m \in \mathcal{I}_{J+1}}, (Z_n^{(m)})_{m \in \mathcal{I}_M \backslash \mathcal{I}_{J+1}}) \in B_n\} \\
= &\lim_{n \to \infty} \Pr\{((X_m^n)_{m \in \mathcal{I}_M}, S^n, f_n^{(J+1)}(X_m^n), (Z_n^{(m)})_{m \in \mathcal{I}_M \backslash \mathcal{I}_{J+1}}) \in C_n\} \\
\ge &\lim_{n \to \infty} \Pr\{((X_m^n)_{m \in \mathcal{I}_M}, S^n, f_n^{(J+1)}(X_m^n), (Z_n^{(m)})_{m \in \mathcal{I}_M \backslash \mathcal{I}_{J+1}}) \in C_n'\} \\
= &1.
\end{IEEEeqnarray*}

Furthermore, Let A be any nonempty subset in $\mathcal{I}_M$. If $J+1 \not \in A$, we have
\begin{IEEEeqnarray*}{rCl}
\underline{I}^{|A|}((\bm{f}^{(m)}(\bm{X}_m))_{m \in A \cap \mathcal{I}_{J+1}}, (\bm{Z}^{(m)})_{m \in A \backslash \mathcal{I}_{J+1}}) &= &\underline{I}^{|A|}((\bm{f}^{(m)}(\bm{X}_m))_{m \in A \cap \mathcal{I}_J}, (\bm{Z}^{(m)})_{m \in A \backslash \mathcal{I}_J}) \\
&\ge &\underline{I}^{|A|}((\bm{Z}^{(m)})_{m \in A}) - \frac{J}{M}\gamma_2 \\
&> &\underline{I}^{|A|}((\bm{Z}^{(m)})_{m \in A}) - \frac{J+1}{M}\gamma_2.
\end{IEEEeqnarray*}
Otherwise,
\begin{IEEEeqnarray*}{Cl}
&\underline{I}^{|A|}((\bm{f}^{(m)}(\bm{X}_m))_{m \in A \cap \mathcal{I}_{J+1}}, (\bm{Z}^{(m)})_{m \in A \backslash \mathcal{I}_{J+1}}) \\
= &\pliminf_{n \to \infty} \frac{1}{n} \ln \frac{P_{(f_n^{(m)}(X_m^n))_{m \in A \cap \mathcal{I}_{J+1}} (Z_n^{(m)})_{m \in A \backslash \mathcal{I}_{J+1}}}((f_n^{(m)}(X_m^n))_{m \in A \cap \mathcal{I}_{J+1}}, (Z_n^{(m)})_{m \in A \backslash \mathcal{I}_{J+1}})}{\prod_{m \in A \cap \mathcal{I}_{J+1}} P_{f_n^{(m)}(X_m^n)}(f_n^{(m)}(X_m^n)) \prod_{m \in A \backslash \mathcal{I}_{J+1}} P_{Z_n^{(m)}}(Z_n^{(m)})} \\
\gevar{(a)} & \pliminf_{n \to \infty} \frac{1}{n} \ln \frac{P_{(f_n^{(m)}(X_m^n))_{m \in A \cap \mathcal{I}_J} (Z_n^{(m)})_{m \in A \backslash \mathcal{I}_J}}((f_n^{(m)}(X_m^n))_{m \in A \cap \mathcal{I}_{J+1}}, (Z_n^{(m)})_{m \in A \backslash \mathcal{I}_{J+1}})}{P_{Z_n^{(J+1)}}(f_n^{(J+1)}(X_{J+1}^{n})) \prod_{m \in A \cap \mathcal{I}_J} P_{f_n^{(m)}(X_m^n)}(f_n^{(m)}(X_m^n)) \prod_{m \in A \backslash \mathcal{I}_{J+1}} P_{Z_n^{(m)}}(Z_n^{(m)})} \\
&+\: \pliminf_{n \to \infty} \frac{1}{n} \ln \frac{P_{(f_n^{(m)}(X_m^n))_{m \in A \cap \mathcal{I}_J} (Z_n^{(m)})_{m \in A \backslash \mathcal{I}_{J+1}} | f_n^{(J+1)}(X_{J+1}^n)}((f_n^{(m)}(X_m^n))_{m \in A \cap \mathcal{I}_J}, (Z_n^{(m)})_{m \in A \backslash \mathcal{I}_{J+1}} | f_n^{(J+1)}(X_{J+1}^n))}{P_{(f_n^{(m)}(X_m^n))_{m \in A \cap \mathcal{I}_J} (Z_n^{(m)})_{m \in A \backslash \mathcal{I}_{J+1}} | Z_n^{(J+1)}}((f_n^{(m)}(X_m^n))_{m \in A \cap \mathcal{I}_J}, (Z_n^{(m)})_{m \in A \backslash \mathcal{I}_{J+1}} | f_n^{(J+1)}(X_{J+1}^{n}))} \\
\gevar{(b)} &\underline{I}^{|A|}((\bm{f}^{(m)}(\bm{X}_m))_{m \in A \cap \mathcal{I}_J}, (\bm{Z}^{(m)})_{m \in A \backslash \mathcal{I}_J}) - \frac{1}{M} \gamma_2 \\
&+\: \underline{D}((f_n^{(m)}(X_m^n))_{m \in A \cap \mathcal{I}_J} (Z_n^{(m)})_{m \in A \backslash \mathcal{I}_{J+1}} | f_n^{(J+1)}(X_{J+1}^n) \| (f_n^{(m)}(X_m^n))_{m \in A \cap \mathcal{I}_J} (Z_n^{(m)})_{m \in A \backslash \mathcal{I}_{J+1}} | Z_n^{(J+1)}) \\
\gevar{(c)} &\underline{I}^{|A|}((\bm{Z}^{(m)})_{m \in A}) - \frac{J+1}{M}\gamma_2,
\end{IEEEeqnarray*}
where (a) follows from the property of the limit inferior in probability, (b) follows from \eqref{eq:DefinitionOfTn1ForLemmaXSZ} and \eqref{eq:C_n'inProbabilityForLemmaXSZ}, and (c) from Lemma \ref{le:InfDx} and \eqref{eq:InfIxForJForLemmaXSZ}.

Analogously, if $J+1 \not \in A$, we have
\begin{IEEEeqnarray*}{Cl}
&\underline{I}((\bm{f}^{(m)}(\bm{X}_m))_{m \in A \cap \mathcal{I}_{J+1}}, (\bm{Z}^{(m)})_{m \in A \backslash \mathcal{I}_{J+1}}; \bm{S}, (\bm{f}^{(m)}(\bm{X}_m))_{m \in (\mathcal{I}_{M} \backslash A) \cap \mathcal{I}_{J+1}}, (\bm{Z}^{(m)})_{m \in (\mathcal{I}_{M} \backslash A) \backslash \mathcal{I}_{J+1}}) \\
= & \underline{I}((\bm{f}^{(m)}(\bm{X}_m))_{m \in A \cap \mathcal{I}_J}, (\bm{Z}^{(m)})_{m \in A \backslash \mathcal{I}_J}; \bm{S}, (\bm{f}^{(m)}(\bm{X}_m))_{m \in (\mathcal{I}_{M} \backslash A) \cap \mathcal{I}_{J+1}}, (\bm{Z}^{(m)})_{m \in (\mathcal{I}_{M} \backslash A) \backslash \mathcal{I}_{J+1}}) \\
= &\pliminf_{n \to \infty} \frac{1}{n} \ln \biggl[ P_{S^n (f_n^{(m)}(X_m^n))_{m \in \mathcal{I}_{J+1}} (Z_n^{(m)})_{m \in \mathcal{I}_M \backslash \mathcal{I}_{J+1}}}(S^n, (f_n^{(m)}(X_m^n))_{m \in \mathcal{I}_{J+1}}, (Z_n^{(m)})_{m \in \mathcal{I}_M \backslash \mathcal{I}_{J+1}}) \bigg/ \\
&\biggl( P_{(f_n^{(m)}(X_m^n))_{m \in A \cap \mathcal{I}_J} (Z_n^{(m)})_{m \in A \backslash \mathcal{I}_J}}((f_n^{(m)}(X_m^n))_{m \in A \cap \mathcal{I}_J}, (Z_n^{(m)})_{m \in A \backslash \mathcal{I}_J}) \\
&P_{S^n (f_n^{(m)}(X_m^n))_{m \in (\mathcal{I}_M \backslash A) \cap \mathcal{I}_{J+1}} (Z_n^{(m)})_{m \in (\mathcal{I}_M \backslash A) \backslash \mathcal{I}_{J+1}}}(S^n, (f_n^{(m)}(X_m^n))_{m \in (\mathcal{I}_M \backslash A) \cap \mathcal{I}_{J+1}}, (Z_n^{(m)})_{m \in (\mathcal{I}_M \backslash A) \backslash \mathcal{I}_{J+1}}) \biggr) \biggr] \\
\gevar{(a)} &\pliminf_{n \to \infty} \frac{1}{n} \ln \biggl( P_{(f_n^{(m)}(X_m^n))_{m \in A \cap \mathcal{I}_J} (Z_n^{(m)})_{m \in A \backslash \mathcal{I}_J} | S^n (f_n^{(m)}(X_m^n))_{m \in (\mathcal{I}_M \backslash A) \cap \mathcal{I}_J} (Z_n^{(m)})_{m \in (\mathcal{I}_M \backslash A) \backslash \mathcal{I}_J}}((f_n^{(m)}(X_m^n))_{m \in A \cap \mathcal{I}_J}, \\
&(Z_n^{(m)})_{m \in A \backslash \mathcal{I}_J} | S^n, (f_n^{(m)}(X_m^n))_{m \in (\mathcal{I}_M \backslash A) \cap \mathcal{I}_{J+1}}, (Z_n^{(m)})_{m \in (\mathcal{I}_M \backslash A) \backslash \mathcal{I}_{J+1}}) \bigg/ \\
&P_{(f_n^{(m)}(X_m^n))_{m \in A \cap \mathcal{I}_J} (Z_n^{(m)})_{m \in A \backslash \mathcal{I}_J}}((f_n^{(m)}(X_m^n))_{m \in A \cap \mathcal{I}_J}, (Z_n^{(m)})_{m \in A \backslash \mathcal{I}_J}) \biggr) \\
&+\: \pliminf_{n \to \infty} \frac{1}{n} \ln \biggl( P_{(f_n^{(m)}(X_m^n))_{m \in A \cap \mathcal{I}_J} (Z_n^{(m)})_{m \in A \backslash \mathcal{I}_J} | S^n (f_n^{(m)}(X_m^n))_{m \in (\mathcal{I}_M \backslash A) \cap \mathcal{I}_{J+1}} (Z_n^{(m)})_{m \in (\mathcal{I}_M \backslash A) \backslash \mathcal{I}_{J+1}}}( \\
&(f_n^{(m)}(X_m^n))_{m \in A \cap \mathcal{I}_J}, (Z_n^{(m)})_{m \in A \backslash \mathcal{I}_J} | S^n, (f_n^{(m)}(X_m^n))_{m \in (\mathcal{I}_M \backslash A) \cap \mathcal{I}_{J+1}}, (Z_n^{(m)})_{m \in (\mathcal{I}_M \backslash A) \backslash \mathcal{I}_{J+1}}) \bigg/ \\
&P_{(f_n^{(m)}(X_m^n))_{m \in A \cap \mathcal{I}_J} (Z_n^{(m)})_{m \in A \backslash \mathcal{I}_J} | S^n (f_n^{(m)}(X_m^n))_{m \in (\mathcal{I}_M \backslash A) \cap \mathcal{I}_J} (Z_n^{(m)})_{m \in (\mathcal{I}_M \backslash A) \backslash \mathcal{I}_J}}((f_n^{(m)}(X_m^n))_{m \in A \cap \mathcal{I}_J}, (Z_n^{(m)})_{m \in A \backslash \mathcal{I}_J} | \\
& S^n, (f_n^{(m)}(X_m^n))_{m \in (\mathcal{I}_M \backslash A) \cap \mathcal{I}_{J+1}}, (Z_n^{(m)})_{m \in (\mathcal{I}_M \backslash A) \backslash \mathcal{I}_{J+1}}) \biggr) \\
\gevar{(b)} &\underline{I}((\bm{f}^{(m)}(\bm{X}_m))_{m \in A \cap \mathcal{I}_J}, (\bm{Z}^{(m)})_{m \in A \backslash \mathcal{I}_J}; \bm{S}, (\bm{f}^{(m)}(\bm{X}_m))_{m \in (\mathcal{I}_{M} \backslash A) \cap \mathcal{I}_J}, (\bm{Z}^{(m)})_{m \in (\mathcal{I}_{M} \backslash A) \backslash \mathcal{I}_J}) - \frac{1}{M} \gamma_2 \\
&+\: \underline{D}((\bm{f}^{(m)}(\bm{X}_m))_{m \in A \cap \mathcal{I}_J}, (\bm{Z}^{(m)})_{m \in A \backslash \mathcal{I}_J} | \bm{S}, (\bm{f}^{(m)}(\bm{X}_m))_{m \in (\mathcal{I}_M \backslash A) \cap \mathcal{I}_{J+1}}, (\bm{Z}^{(m)})_{m \in (\mathcal{I}_M \backslash A) \backslash \mathcal{I}_{J+1}} \| \\
&(\bm{f}^{(m)}(\bm{X}_m))_{m \in A \cap \mathcal{I}_J}, (\bm{Z}^{(m)})_{m \in A \backslash \mathcal{I}_J} | \bm{S}, (\bm{f}^{(m)}(\bm{X}_m))_{m \in (\mathcal{I}_M \backslash A) \cap \mathcal{I}_J}, (\bm{Z}^{(m)})_{m \in (\mathcal{I}_M \backslash A) \backslash \mathcal{I}_J}) \\
\gevar{(c)} &\underline{I}((\bm{Z}^{(m)})_{m \in A}; \bm{S}, (\bm{Z}^{(m)})_{m \in \mathcal{I}_{M} \backslash A}) - \frac{J+1}{M}\gamma_2,
\end{IEEEeqnarray*}
where (a) follows from the property of the limit inferior in probability, (b) follows from \eqref{eq:DefinitionOfTn2ForLemmaXSZ} and \eqref{eq:C_n'inProbabilityForLemmaXSZ}, and (c) from Lemma \ref{le:InfDx} and \eqref{eq:InfIForLemmaXSZ}. In the same way, it can also be shown that
\begin{IEEEeqnarray*}{Cl}
&\underline{I}((\bm{f}^{(m)}(\bm{X}_m))_{m \in A \cap \mathcal{I}_{J+1}}, (\bm{Z}^{(m)})_{m \in A \backslash \mathcal{I}_{J+1}}; \bm{S}, (\bm{f}^{(m)}(\bm{X}_m))_{m \in (\mathcal{I}_{M} \backslash A) \cap \mathcal{I}_{J+1}}, (\bm{Z}^{(m)})_{m \in (\mathcal{I}_{M} \backslash A) \backslash \mathcal{I}_{J+1}}) \\
\ge &\underline{I}((\bm{Z}^{(m)})_{m \in A}; \bm{S}, (\bm{Z}^{(m)})_{m \in \mathcal{I}_{M} \backslash A}) - \frac{J+1}{M}\gamma_2
\end{IEEEeqnarray*}
for the case of $J+1 \in A$.

Therefore, the conditions \eqref{eq:fnForLemmaXSZ}, \eqref{eq:BnInProbabilityForLemmaXSZ}, \eqref{eq:InfIxForJForLemmaXSZ} and \eqref{eq:InfIForLemmaXSZ} hold for $J+1$. The lemma is hence proved by simply repeating the above argument $M$ times.
\end{proof}

Now, we start to prove Theorem \ref{th:AchievableCondition}.

\begin{proofof}{Theorem \ref{th:AchievableCondition}}
1) \textit{Direct Part:} Let $\gamma_1$, $\gamma_2$, $\gamma_3$ and $\gamma_4$ be arbitrary positive real numbers. We define $T_n^{(1)}$ by
\begin{IEEEeqnarray*}{rCl}
T_n^{(1)} &\eqdef &\biggl\{((\bm{x}_m)_{m \in \mathcal{I}_M}, \bm{s}, (\bm{z}^{(m)})_{m \in \mathcal{I}_M}) \in \prod_{m \in \mathcal{I}_M} \mathcal{X}_m^n \times \mathcal{S}^n \times \prod_{m \in \mathcal{I}_M} \mathcal{Z}_n^{(m)} \bigg| \\
& &(d_n^{(k)}((\bm{x}_m)_{m \in \mathcal{I}_{M}}, h_n(\bm{s}, (\bm{z}^{(m)})_{m \in \mathcal{I}_{M}})))_{k \in \mathcal{I}_{K}} \le (D_k + \gamma_1)_{k \in \mathcal{I}_{K}} \biggr\}.
\end{IEEEeqnarray*}
Clearly, by the condition \eqref{eq:AchievableCondition2}, we have
$$
\lim_{n \to \infty} \Pr\{((X_m^n)_{m \in \mathcal{I}_M}, S^n, (Z_n^{(m)})_{m \in \mathcal{I}_M}) \in T_n^{(1)}\} = 1.
$$
Then by Lemma \ref{le:XSZ}, we obtain a group of sequences $(\bm{f}^{(m)} = \{f_n^{(m)}\}_{n=1}^\infty)_{m \in \mathcal{I}_M}$ of functions $f_n^{(m)}: \mathcal{X}_m^n \to \mathcal{Z}_n^{(m)}$ such that
$$
|f_n^{(m)}(\mathcal{X}_m^n)| \le e^{\ceil{n(\overline{I}(\bm{X}_m; \bm{Z}^{(m)}) + \gamma_2)}},
$$
\begin{equation}\label{eq:AlmostSureTn1}
\lim_{n \to \infty} \Pr\{((X_m^n)_{m \in \mathcal{I}_M}, S^n, (f_n^{(m)}(X_m^n))_{m \in \mathcal{I}_M}) \in T_n^{(1)}\} = 1,
\end{equation}
and
\begin{equation}\label{eq:Property1FromLemmaXYZ}
\underline{I}^{|A|}((\bm{f}^{(m)}(\bm{X}_m))_{m \in A}) \ge \underline{I}^{|A|}((\bm{Z}^{(m)})_{m \in A}) - \gamma_3,
\end{equation}
\begin{equation}\label{eq:Property2FromLemmaXYZ}
\underline{I}((\bm{f}^{(m)}(\bm{X}_m))_{m \in A}; \bm{S}, (\bm{f}^{(m)}(\bm{X}_m))_{m \in \mathcal{I}_{M} \backslash A}) \ge \underline{I}((\bm{Z}^{(m)})_{m \in A}; \bm{S}, (\bm{Z}^{(m)})_{m \in \mathcal{I}_{M} \backslash A}) - \gamma_3
\end{equation}
for any nonempty set $A \subseteq \mathcal{I}_{M}$, where $\bm{f}^{(m)}(\bm{X}_m) \eqdef \{f_n^{(m)}(X_m^n)\}_{n=1}^\infty$.

Next, we specify the encoding and decoding procedures. We will present a pair of random encoding and decoding maps according to a given rate $(R_m)_{m \in \mathcal{I}_M}$ satisfying
\begin{equation}
\sum_{m \in A} R_m \ge \sum_{m \in A} \overline{I}(\bm{X}_m; \bm{Z}^{(m)}) - \underline{I}^{|A|}((\bm{Z}^{(m)})_{m \in A}) - \underline{I}((\bm{Z}^{(m)})_{m \in A}; \bm{S}, (\bm{Z}^{(m)})_{m \in \mathcal{I}_{M} \backslash A})
\end{equation}
for any nonempty set $A \subseteq \mathcal{I}_{M}$.

\textit{Generation of random bins:} For each $m \in \mathcal{I}_{M}$, assign each $\bm{z}^{(m)} \in \mathcal{Z}_n^{(m)}$ to one of the indices in $\mathcal{I}_{L_n^{(m)}}$ according to a uniform distribution on $\mathcal{I}_{L_n^{(m)}}$ independently, where
\begin{equation}\label{eq:RateDefinition}
L_n^{(m)} \eqdef \ceil{e^{n(R_m+\gamma_1)}}.
\end{equation}
Let $G_n^{(m)}(\bm{z}^{(m)})$ denote the index to which $\bm{z}^{(m)}$ corresponds.

\textit{Encoding} $(\Phi_n^{(m)}: \mathcal{X}_m^n \to \mathcal{I}_{L_n^{(m)}})_{m \in \mathcal{I}_M}$. For each $m \in \mathcal{I}_M$, the encoder $\Phi_n^{(m)}$ with respect to the $m$-th terminal is defined by
$$
\Phi_n^{(m)}(\bm{x}_m) \eqdef G_n^{(m)}(f_n^{(m)}(\bm{x}_m))
$$
for all $\bm{x}_m \in \mathcal{X}_m^n$.

\textit{Decoding} $\Psi_n: \mathcal{S}^n \times \prod_{m \in \mathcal{I}_M} \mathcal{I}_{L_n^{(m)}} \to \prod_{m \in \mathcal{I}_M} \mathcal{Y}_m^n$. The decoder receives the pair $(\bm{s}, (\Phi_n^{(m)}(\bm{x}_m))_{m \in \mathcal{I}_M}) \in \mathcal{S}^n \times \prod_{m \in \mathcal{I}_M} \mathcal{I}_{L_n^{(m)}}$. For $(\bm{s}, (\Phi_n^{(m)}(\bm{x}_m))_{m \in \mathcal{I}_M})$, if there is one and only one $(\bm{z}^{(m)})_{m \in \mathcal{I}_M} \in \prod_{m \in \mathcal{I}_M} \mathcal{Z}_n^{(m)}$ satisfying
$$
(G_n^{(m)}(\bm{z}^{(m)}))_{m \in \mathcal{I}_M} = (\Phi_n^{(m)}(\bm{x}_m))_{m \in \mathcal{I}_M}
$$
and
$$
(\bm{s}, (\bm{z}^{(m)})_{m \in \mathcal{I}_M}) \in T_n^{(2)},
$$
we declare $\Psi_n(\bm{s}, (\Phi_n^{(m)}(\bm{x}_m))_{m \in \mathcal{I}_M}) = h_n(\bm{s}, (\bm{z}^{(m)})_{m \in \mathcal{I}_M})$, where
\begin{equation}\label{eq:DefinitionOfTn2}
T_n^{(2)} \eqdef T_n^{(2, 0)} \cap \bigcap_{B \subseteq \mathcal{I}_M, B \neq \emptyset} T_n^{(2, A)},
\end{equation}
$$
T_n^{(2, 0)} \eqdef \biggl\{(\bm{s}, (\bm{z}^{(m)})_{m \in \mathcal{I}_M}) \in \mathcal{S}^n \times \prod_{m \in \mathcal{I}_M} \mathcal{Z}_n^{(m)} \bigg| \frac{1}{n} \ln \frac{1}{P_{f_n^{(m)}(X_m^n)}(\bm{z}^{(m)})} \le \overline{I}(\bm{X}_m; \bm{Z}^{(m)}) + 2\gamma_2, \forall m \in \mathcal{I}_M \biggr\},
$$
\begin{IEEEeqnarray*}{rCl}
T_n^{(2, B)} &\eqdef &\biggl\{(\bm{s}, (\bm{z}^{(m)})_{m \in \mathcal{I}_M}) \in \mathcal{S}^n \times \prod_{m \in \mathcal{I}_M} \mathcal{Z}_n^{(m)} \bigg| \frac{1}{n} \ln \frac{P_{(f_n^{(m)}(X_m^n))_{m \in B}}((\bm{z}^{(m)})_{m \in B})}{\prod_{m \in B} P_{f_n^{(m)}(X_m^n)}(\bm{z}^{(m)})} \ge \underline{I}^{|B|}((\bm{f}^{(m)}(\bm{X}_m))_{m \in B}) - \gamma_4, \\
& &\frac{1}{n} \ln \frac{P_{S^n (f_n^{(m)}(X_m^n))_{m \in \mathcal{I}_M}}(\bm{s}, (\bm{z}^{(m)})_{m \in \mathcal{I}_M})}{P_{(f_n^{(m)}(X_m^n))_{m \in B}}((\bm{z}^{(m)})_{m \in B}) P_{S^n (f_n^{(m)}(X_m^n))_{m \in \mathcal{I}_M \backslash B}}(\bm{s}, (\bm{z}^{(m)})_{m \in \mathcal{I}_M \backslash B})} \\
& &\ge \underline{I}((\bm{f}^{(m)}(\bm{X}_m))_{m \in B}; \bm{S}, (\bm{f}^{(m)}(\bm{X}_m))_{m \in \mathcal{I}_{M} \backslash B}) - \gamma_4 \biggr\}.
\end{IEEEeqnarray*}
Otherwise, $\Psi_n(\bm{s}, (\Phi_n^{(m)}(\bm{x}_m))_{m \in \mathcal{I}_M})$ is defined to be an arbitrary fixed element in $\prod_{m \in \mathcal{I}_M} \mathcal{Y}_m^n$.

If the source outputs $(\bm{x}_m)_{m \in \mathcal{I}_M}$ of the $m$ terminals together with the side information $\bm{s}$ satisfy the following conditions:

(1) $((\bm{x}_m)_{m \in \mathcal{I}_M}, \bm{s}, (f_n^{(m)}(\bm{x}_m))_{m \in \mathcal{I}_M}) \in T_n^{(1)} \cap (\prod_{m \in \mathcal{I}_M} \mathcal{X}_m^n \times T_n^{(2)})$\par
(2) There is no $(\bm{z}^{(m)})_{m \in \mathcal{I}_M} \in \prod_{m \in \mathcal{I}_M} \mathcal{Z}_n^{(m)}$ such that
$$
(\bm{z}^{(m)})_{m \in \mathcal{I}_M} \neq (f_n^{(m)}(\bm{x}_m))_{m \in \mathcal{I}_M},
$$
$$
(G_n^{(m)}(\bm{z}^{(m)}))_{m \in \mathcal{I}_M} = (\Phi_n^{(m)}(\bm{x}_m))_{m \in \mathcal{I}_M},
$$
$$
(\bm{s}, (\bm{z}^{(m)})_{m \in \mathcal{I}_M}) \in T_n^{(2)},
$$
then $(f_n^{(m)}(\bm{x}_m))_{m \in \mathcal{I}_M}$ is the unique element in $\prod_{m \in \mathcal{I}_M} \mathcal{Z}_n^{(m)}$ such that
$$
(G_n^{(m)}(f_n^{(m)}(\bm{x}_m)))_{m \in \mathcal{I}_M} = (\Phi_n^{(m)}(\bm{x}_m))_{m \in \mathcal{I}_M}
$$
and
$$
(\bm{s}, (f_n^{(m)}(\bm{x}_m))_{m \in \mathcal{I}_M}) \in T_n^{(2)}.
$$
Furthermore, we have
$$
d_n^{(k)}((\bm{x}_m)_{m \in \mathcal{I}_M}, \Psi_n(\bm{s}, (\Phi_n^{(m)}(\bm{x}_m))_{m \in \mathcal{I}_M})) = d_n^{(k)}((\bm{x}_m)_{m \in \mathcal{I}_M}, h_n(\bm{s}, (f_n^{(m)}(\bm{x}_m))_{m \in \mathcal{I}_M})) \le D_k + \gamma_1, \quad \forall k \in \mathcal{I}_K.
$$
due to $((\bm{x}_m)_{m \in \mathcal{I}_M}, \bm{s}, (f_n^{(m)}(\bm{x}_m))_{m \in \mathcal{I}_M}) \in T_n^{(1)}$. Hence, every pair $((\bm{x}_m)_{m \in \mathcal{I}_M}, \bm{s}) \in \mathcal{T}_1(\Phi_n, \Psi_n) \cap \mathcal{T}_2(\Phi_n, \Psi_n)$ satisfies
$$
(d_n^{(k)}((\bm{x}_m)_{m \in \mathcal{I}_M}, \Psi_n(\bm{s}, (\Phi_n^{(m)}(\bm{x}_m))_{m \in \mathcal{I}_M})))_{k \in \mathcal{I}_{K}} \le (D_k + \gamma_1)_{k \in \mathcal{I}_{K}},
$$
where $\mathcal{T}_i(\Phi_n, \Psi_n)$ is defined for the code $(\Phi_n, \Psi_n)$ by
$$
\mathcal{T}_i(\Phi_n, \Psi_n) \eqdef \biggl\{ ((\bm{x}_m)_{m \in \mathcal{I}_M}, \bm{s}) \in \prod_{m \in \mathcal{I}_M} \mathcal{X}_m^n \times \mathcal{S}^n \bigg| ((\bm{x}_m)_{m \in \mathcal{I}_M}, \bm{s}) \mbox{ satisfies condition $i$} \biggr\}
$$
for $i = 1, 2$.

Next, we estimate the probability of error. Let us define the probability of error $P_e^{(n)}(\Phi_n, \Psi_n)$ for the code $(\Phi_n, \Psi_n)$ by
$$
P_e^{(n)}(\Phi_n, \Psi_n) = \Pr\{\exists k \in \mathcal{I}_K, \suchthat d_n^{(k)}((X_m^n)_{m \in \mathcal{I}_M}, \Psi_n(S^n, (\Phi_n^{(m)}(X_m^n))_{m \in \mathcal{I}_M})) > D_k + \gamma_1\}.
$$
Then, the $P_e^{(n)}(\Phi_n, \Psi_n)$ can be bounded as follows:
\begin{IEEEeqnarray}{rCl}
P_e^{(n)}(\Phi_n, \Psi_n) &\le &1 - \Pr\{((X_m^n)_{m \in \mathcal{I}_M}, S^n) \in \mathcal{T}_1(\Phi_n, \Psi_n) \cap \mathcal{T}_2(\Phi_n, \Psi_n)\} \IEEEnonumber \\
&= &\Pr\{((X_m^n)_{m \in \mathcal{I}_M}, S^n) \not \in \mathcal{T}_1(\Phi_n, \Psi_n) \cap \mathcal{T}_2(\Phi_n, \Psi_n)\} \IEEEnonumber \\
&\le &\Pr\{((X_m^n)_{m \in \mathcal{I}_M}, S^n) \not \in \mathcal{T}_1(\Phi_n, \Psi_n)\} + \Pr\{((X_m^n)_{m \in \mathcal{I}_M}, S^n) \not \in \mathcal{T}_2(\Phi_n, \Psi_n)\}.\label{eq:UpperBoundOnPe}
\end{IEEEeqnarray}
Clearly, the first term in \eqref{eq:UpperBoundOnPe} can be bounded above by
$$
\Pr\{((X_m^n)_{m \in \mathcal{I}_M}, S^n, (f_n^{(m)}(X_m^n))_{m \in \mathcal{I}_M}) \not \in T_n^{(1)}\} + \Pr\{(S^n, (f_n^{(m)}(X_m^n))_{m \in \mathcal{I}_M}) \not \in T_n^{(2)}\}
$$
and hence converges to zero by \eqref{eq:AlmostSureTn1} and \eqref{eq:DefinitionOfTn2}. As for the second term in \eqref{eq:UpperBoundOnPe}, the event $((X_m^n)_{m \in \mathcal{I}_M}, S^n) \not \in \mathcal{T}_2(\Phi_n, \Psi_n)$ can be rewritten as
$$
((X_m^n)_{m \in \mathcal{I}_M}, S^n) \in \bigcup_{B \subseteq \mathcal{I}_M, B \neq \emptyset} E_n(\Phi_n, \Psi_n, B),
$$
where
\begin{IEEEeqnarray*}{rCl}
E_n(\Phi_n, \Psi_n, B) &= &\{((\bm{x}_m)_{m \in \mathcal{I}_M}, \bm{s}) \in \prod_{m \in \mathcal{I}_M} \mathcal{X}_m^n \times \mathcal{S}^n | \exists (\bm{z}^{(m)})_{m \in B} \in \prod_{m \in B} \mathcal{Z}_n^{(m)}, \suchthat (\bm{z}^{(m)})_{m \in B} \neq (f_n^{(m)}(\bm{x}_m))_{m \in B}, \\
& &(G_n^{(m)}(\bm{z}^{(m)}))_{m \in B} = (\Phi_n^{(m)}(\bm{x}_m))_{m \in B}, (\bm{s}, (\bm{z}^{(m)})_{m \in B}, (f_n^{(m)}(\bm{x}_m))_{m \in \mathcal{I}_M \backslash B}) \in T_n^{(2)}\}.
\end{IEEEeqnarray*}
Then we have
\begin{equation}\label{eq:SummationForErrorOfCondition2}
\Pr\{((X_m^n)_{m \in \mathcal{I}_M}, S^n) \not \in \mathcal{T}_2(\Phi_n, \Psi_n)\} \le \sum_{B \subseteq \mathcal{I}_M, B \neq \emptyset} \Pr\{((X_m^n)_{m \in \mathcal{I}_M}, S^n) \in E_n(\Phi_n, \Psi_n, B)\}
\end{equation}
In order to estimate $\Pr\{((X_m^n)_{m \in \mathcal{I}_M}, S^n) \not \in \mathcal{T}_2(\Phi_n, \Psi_n)\}$, let us estimate each term in the summation \eqref{eq:SummationForErrorOfCondition2}. For each nonempty set $B \subseteq \mathcal{I}_M$, we have
\begin{IEEEeqnarray*}{Cl}
&\Pr\{((X_m^n)_{m \in \mathcal{I}_M}, S^n) \in E_n(\Phi_n, \Psi_n, B)\} \\
= &\sum_{g_n} P_{G_n}(g_n) \sum_{((\bm{x}_m)_{m \in \mathcal{I}_M}, \bm{s}) \in \prod_{m \in \mathcal{I}_M} \mathcal{X}_m^n \times \mathcal{S}^n} P_{(X_m^n)_{m \in \mathcal{I}_M}S^n}((\bm{x}_m)_{m \in \mathcal{I}_M}, \bm{s}) 1\{\exists (\bm{z}^{(m)})_{m \in B} \in \prod_{m \in B} \mathcal{Z}_n^{(m)}, \suchthat (\bm{z}^{(m)})_{m \in B} \\
&\neq (f_n^{(m)}(\bm{x}_m))_{m \in B}, (g_n^{(m)}(\bm{z}^{(m)}))_{m \in B} = (g_n^{(m)}(f_n^{(m)}(\bm{x}_m)))_{m \in B}, (\bm{s}, (\bm{z}^{(m)})_{m \in B}, (f_n^{(m)}(\bm{x}_m))_{m \in \mathcal{I}_M \backslash B}) \in T_n^{(2)}\} \\
\le &\sum_{g_n} P_{G_n}(g_n) \sum_{((\bm{x}_m)_{m \in \mathcal{I}_M}, \bm{s}) \in \prod_{m \in \mathcal{I}_M} \mathcal{X}_m^n \times \mathcal{S}^n} P_{(X_m^n)_{m \in \mathcal{I}_M}S^n}((\bm{x}_m)_{m \in \mathcal{I}_M}, \bm{s}) \\
&\sum_{\scriptstyle (\bm{z}^{(m)})_{m \in B} \in \prod_{m \in B} \mathcal{Z}_n^{(m)}, (\bm{z}^{(m)})_{m \in B} \neq (f_n^{(m)}(\bm{x}_m))_{m \in B}, \atop \scriptstyle (\bm{s}, (\bm{z}^{(m)})_{m \in B}, (f_n^{(m)}(\bm{x}_m))_{m \in \mathcal{I}_M \backslash B}) \in T_n^{(2)}} 1\{(g_n^{(m)}(\bm{z}^{(m)}))_{m \in B} = (g_n^{(m)}(f_n^{(m)}(\bm{x}_m)))_{m \in B}\} \\
= &\sum_{((\bm{x}_m)_{m \in \mathcal{I}_M}, \bm{s}) \in \prod_{m \in \mathcal{I}_M} \mathcal{X}_m^n \times \mathcal{S}^n} P_{(X_m^n)_{m \in \mathcal{I}_M}S^n}((\bm{x}_m)_{m \in \mathcal{I}_M}, \bm{s}) \sum_{\scriptstyle (\bm{z}^{(m)})_{m \in B} \in \prod_{m \in B} \mathcal{Z}_n^{(m)}, (\bm{z}^{(m)})_{m \in B} \neq (f_n^{(m)}(\bm{x}_m))_{m \in B}, \atop \scriptstyle (\bm{s}, (\bm{z}^{(m)})_{m \in B}, (f_n^{(m)}(\bm{x}_m))_{m \in \mathcal{I}_M \backslash B}) \in T_n^{(2)}} \\
&\Pr\{(G_n^{(m)}(\bm{z}^{(m)}))_{m \in B} = (G_n^{(m)}(f_n^{(m)}(\bm{x}_m)))_{m \in B}\} \\
\eqvar{(a)} &\sum_{((\bm{x}_m)_{m \in \mathcal{I}_M}, \bm{s}) \in \prod_{m \in \mathcal{I}_M} \mathcal{X}_m^n \times \mathcal{S}^n} P_{(X_m^n)_{m \in \mathcal{I}_M}S^n}((\bm{x}_m)_{m \in \mathcal{I}_M}, \bm{s}) \sum_{\scriptstyle (\bm{z}^{(m)})_{m \in B} \in \prod_{m \in B} \mathcal{Z}_n^{(m)}, (\bm{z}^{(m)})_{m \in B} \neq (f_n^{(m)}(\bm{x}_m))_{m \in B}, \atop \scriptstyle (\bm{s}, (\bm{z}^{(m)})_{m \in B}, (f_n^{(m)}(\bm{x}_m))_{m \in \mathcal{I}_M \backslash B}) \in T_n^{(2)}} \frac{1}{\prod_{m \in B} L_n^{(m)}} \\
\le &\sum_{((\bm{x}_m)_{m \in \mathcal{I}_M}, \bm{s}) \in \prod_{m \in \mathcal{I}_M} \mathcal{X}_m^n \times \mathcal{S}^n} P_{(X_m^n)_{m \in \mathcal{I}_M}S^n}((\bm{x}_m)_{m \in \mathcal{I}_M}, \bm{s}) \sum_{\scriptstyle (\bm{z}^{(m)})_{m \in B} \in \prod_{m \in B} \mathcal{Z}_n^{(m)}, \atop \scriptstyle (\bm{s}, (\bm{z}^{(m)})_{m \in B}, (f_n^{(m)}(\bm{x}_m))_{m \in \mathcal{I}_M \backslash B}) \in T_n^{(2)}} \frac{1}{\prod_{m \in B} L_n^{(m)}} \\
= &\sum_{((\bm{x}_m)_{m \in \mathcal{I}_M \backslash B}, \bm{s}) \in \prod_{m \in \mathcal{I}_M \backslash B} \mathcal{X}_m^n \times \mathcal{S}^n} P_{(X_m^n)_{m \in \mathcal{I}_M \backslash B}S^n}((\bm{x}_m)_{m \in \mathcal{I}_M \backslash B}, \bm{s}) \sum_{\scriptstyle (\bm{z}^{(m)})_{m \in B} \in \prod_{m \in B} \mathcal{Z}_n^{(m)}, \atop \scriptstyle (\bm{s}, (\bm{z}^{(m)})_{m \in B}, (f_n^{(m)}(\bm{x}_m))_{m \in \mathcal{I}_M \backslash B}) \in T_n^{(2)}} \frac{1}{\prod_{m \in B} L_n^{(m)}} \\
= &\sum_{(\bm{s}, (\bm{z}^{(m)})_{m \in \mathcal{I}_M \backslash B}) \in \mathcal{S}^n \times \prod_{m \in \mathcal{I}_M \backslash B} \mathcal{Z}_n^{(m)}} P_{S^n(f_n^{(m)}(X_m^n))_{m \in \mathcal{I}_M \backslash B}}(\bm{s}, (\bm{z}^{(m)})_{m \in \mathcal{I}_M \backslash B}) \sum_{\scriptstyle (\bm{z}^{(m)})_{m \in B} \in \prod_{m \in B} \mathcal{Z}_n^{(m)}, \atop \scriptstyle (\bm{s}, (\bm{z}^{(m)})_{m \in \mathcal{I}_M}) \in T_n^{(2)}} \frac{1}{\prod_{m \in B} L_n^{(m)}} \\
= &\sum_{(\bm{s}, (\bm{z}^{(m)})_{m \in \mathcal{I}_M}) \in T_n^{(2)}} \frac{P_{S^n(f_n^{(m)}(X_m^n))_{m \in \mathcal{I}_M \backslash B}}(\bm{s}, (\bm{z}^{(m)})_{m \in \mathcal{I}_M \backslash B})}{\prod_{m \in B} L_n^{(m)}} \\
\levar{(b)} &\sum_{(\bm{s}, (\bm{z}^{(m)})_{m \in \mathcal{I}_M}) \in T_n^{(2, 0)} \cap T_n^{(2, B)}} \frac{P_{S^n(f_n^{(m)}(X_m^n))_{m \in \mathcal{I}_M \backslash B}}(\bm{s}, (\bm{z}^{(m)})_{m \in \mathcal{I}_M \backslash B})}{\prod_{m \in B} L_n^{(m)}} \\
\levar{(c)} &\sum_{(\bm{z}^{(m)})_{m \in B} \in \prod_{m \in B} \mathcal{Z}_n^{(m)}} 1\biggl\{\frac{1}{n} \ln \frac{P_{(f_n^{(m)}(X_m^n))_{m \in B}}((\bm{z}^{(m)})_{m \in B})}{\prod_{m \in B} P_{f_n^{(m)}(X_m^n)}(\bm{z}^{(m)})} \ge \underline{I}^{|B|}((\bm{f}^{(m)}(\bm{X}_m))_{m \in B}) - \gamma_4\biggr\} \\
&1\biggl\{ \frac{1}{n} \ln \frac{1}{P_{f_n^{(m)}(X_m^n)}(\bm{z}^{(m)})} \le \overline{I}(\bm{X}_m; \bm{Z}^{(m)}) + 2\gamma_2, \forall m \in B \biggr\} \\
&\sum_{(\bm{s}, (\bm{z}^{(m)})_{m \in \mathcal{I}_M \backslash B}) \in \mathcal{S}^n \times \prod_{m \in \mathcal{I}_M \backslash B} \mathcal{Z}_n^{(m)}} \frac{P_{S^n(f_n^{(m)}(X_m^n))_{m \in \mathcal{I}_M \backslash B} | (f_n^{(m)}(X_m^n))_{m \in B}}(\bm{s}, (\bm{z}^{(m)})_{m \in \mathcal{I}_M \backslash B} | (\bm{z}^{(m)})_{m \in B})}{e^{n(\underline{I}((\bm{f}^{(m)}(\bm{X}_m))_{m \in B}; \bm{S}, (\bm{f}^{(m)}(\bm{X}_m))_{m \in \mathcal{I}_{M} \backslash B}) - \gamma_4)} \prod_{m \in B} L_n^{(m)}} \\
\le &\frac{\sum_{(\bm{z}^{(m)})_{m \in B} \in \prod_{m \in B} \mathcal{Z}_n^{(m)}} 1\biggl\{P_{(f_n^{(m)}(X_m^n))_{m \in B}}((z_n^{(m)})_{m \in B}) \ge e^{n(\underline{I}^{|B|}((\bm{f}^{(m)}(\bm{X}_m))_{m \in B}) - \sum_{m \in B} \overline{I}(\bm{X}_m; \bm{Z}^{(m)}) - 2|B|\gamma_2 - \gamma_4)} \biggr\}}{e^{n(\underline{I}((\bm{f}^{(m)}(\bm{X}_m))_{m \in B}; \bm{S}, (\bm{f}^{(m)}(\bm{X}_m))_{m \in \mathcal{I}_{M} \backslash B}) - \gamma_4)} \prod_{m \in B} L_n^{(m)}} \\
\le &e^{-n(\underline{I}^{|B|}((\bm{f}^{(m)}(\bm{X}_m))_{m \in B}) + \underline{I}((\bm{f}^{(m)}(\bm{X}_m))_{m \in B}; \bm{S}, (\bm{f}^{(m)}(\bm{X}_m))_{m \in \mathcal{I}_{M} \backslash B}) - \sum_{m \in B} \overline{I}(\bm{X}_m; \bm{Z}^{(m)}) - 2|B|\gamma_2 - 2 \gamma_4)} \biggl( \prod_{m \in B} L_n^{(m)} \biggr)^{-1} \\
\levar{(d)} &e^{-n[\sum_{m \in B} R_m - (\sum_{m \in B} \overline{I}(\bm{X}_m; \bm{Z}^{(m)}) - \underline{I}^{|B|}((\bm{Z}^{(m)})_{m \in B}) - \underline{I}((\bm{Z}^{(m)})_{m \in B}; \bm{S}, (\bm{Z}^{(m)})_{m \in \mathcal{I}_{M} \backslash B})) + |B|(\gamma_1 - 2\gamma_2) - 2(\gamma_3 + \gamma_4)]}
\end{IEEEeqnarray*}
where (a) follows from the property of the random bins, (b) and (c) from the definition \eqref{eq:DefinitionOfTn2}, and (d) from \eqref{eq:Property1FromLemmaXYZ}, \eqref{eq:Property2FromLemmaXYZ} and \eqref{eq:RateDefinition}.

Since $\gamma_2$, $\gamma_3$ and $\gamma_4$ are arbitrary, let us define
$$
\gamma_2 = \gamma_3 = \gamma_4 < \frac{\gamma_1}{6},
$$
then
$$
\lim_{n \to \infty} \Pr\{((X_m^n)_{m \in \mathcal{I}_M}, S^n) \in E_n(\Phi_n, \Psi_n, B)\} = 0
$$
for all nonempty set $B \subseteq \mathcal{I}_M$, and hence we have
\begin{IEEEeqnarray*}{rCl}
\limsup_{n \to \infty} P_e^{(n)}(\Phi_n, \Psi_n) &\le &\limsup_{n \to \infty} \biggl( \Pr\{((X_m^n)_{m \in \mathcal{I}_M}, S^n) \not \in \mathcal{T}_1(\Phi_n, \Psi_n)\} + \Pr\{((X_m^n)_{m \in \mathcal{I}_M}, S^n) \not \in \mathcal{T}_2(\Phi_n, \Psi_n)\} \biggr) \\
&\le &\limsup_{n \to \infty} \Pr\{((X_m^n)_{m \in \mathcal{I}_M}, S^n) \not \in \mathcal{T}_1(\Phi_n, \Psi_n)\} + \limsup_{n \to \infty} \Pr\{((X_m^n)_{m \in \mathcal{I}_M}, S^n) \not \in \mathcal{T}_2(\Phi_n, \Psi_n)\} \\
&\le &\sum_{B \subseteq \mathcal{I}_M, B \neq \emptyset} \limsup_{n \to \infty} \Pr\{((X_m^n)_{m \in \mathcal{I}_M}, S^n) \in E_n(\Phi_n, \Psi_n, B)\} \\
&= &0,
\end{IEEEeqnarray*}
which implies that there exists at least one sequence $\{(\phi_n, \psi_n)\}_{n=1}^{\infty}$ of codes satisfying
$$
\plimsup_{n \to \infty} d_n^{(k)}((X_m^n)_{m \in \mathcal{I}_M}, \psi_n(S^n, (\phi_n^{(m)}(X_m^n))_{m \in \mathcal{I}_M})) \le D_k + \gamma_1, \quad \forall k \in \mathcal{I}_K
$$
and
$$
\limsup_{n \to \infty} R(\phi_n^{(m)}) = R_m + \gamma_1, \quad \forall m \in \mathcal{I}_M.
$$

Finally, let us use the diagonal method to complete the proof. By repeating the above argument with replacing $\gamma_1$ by a sequence $\{\gamma_1(i)\}_{i=1}^\infty$ which satisfies $\gamma_1(1) \ge \gamma_1(2) \ge \cdots > 0$ and $\gamma_1(i) \to 0$ as $i \to \infty$, we can conclude that there exists a sequence $\{(\phi_n, \psi_n)\}_{n=1}^\infty$ of codes which satisfies
$$
(\limsup_{n \to \infty} R(\phi_n^{(m)}))_{m \in \mathcal{I}_{M}} \le (R_m)_{m \in \mathcal{I}_{M}},
$$
$$
(\plimsup_{n \to \infty} d_n^{(k)}((X_m^n)_{m \in \mathcal{I}_{M}}, \psi_n(S^n, (\phi_n^{(m)}(X_m^n))_{m \in \mathcal{I}_{M}})))_{k \in \mathcal{I}_{K}} \le (D_k)_{k \in \mathcal{I}_{K}}.
$$
Therefore, the proof of the direct part is established.

2) Converse Part: Since the rate pair $((R_m)_{m \in \mathcal{I}_M}, (D_k)_{k \in \mathcal{I}_K}$ is achievable, there exits a sequence $\{(\phi_n, \psi_n)\}_{n=1}^\infty$ such that
$$
(\limsup_{n \to \infty} R(\phi_n^{(m)}))_{m \in \mathcal{I}_{M}} \le (R_m)_{m \in \mathcal{I}_{M}},
$$
$$
(\plimsup_{n \to \infty} d_n^{(k)}((X_m^n)_{m \in \mathcal{I}_{M}}, \psi_n(S^n, (\phi_n^{(m)}(X_m^n))_{m \in \mathcal{I}_{M}})))_{k \in \mathcal{I}_{K}} \le (D_k)_{k \in \mathcal{I}_{K}}.
$$
Let us define the general sources $(\bm{Z}^{(m)})_{m \in \mathcal{I}_M}$ by $Z_n^{(m)} = \phi_n(X_m^n)$ ($m \in \mathcal{I}_M$), then
$$
P_{(X_m^n)_{m \in \mathcal{I}_{M}} S^n (Z_n^{(m)})_{m \in \mathcal{I}_{M}}} = P_{(X_m^n)_{m \in \mathcal{I}_{M}} S^n} \prod_{m \in \mathcal{I}_{M}} P_{Z_n^{(m)} | X_m^n}
$$
holds for all $n = 1, 2, \cdots$. Besides, let us define $h_n = \psi_n$, then we have
\begin{IEEEeqnarray*}{Cl}
&(\plimsup_{n \to \infty} d_n^{(k)}((X_m^n)_{m \in \mathcal{I}_{M}}, h_n(S^n, (Z_n^{(m)})_{m \in \mathcal{I}_{M}})))_{k \in \mathcal{I}_{K}} \\
= &(\plimsup_{n \to \infty} d_n^{(k)}((X_m^n)_{m \in \mathcal{I}_{M}}, \psi_n(S^n, (\phi_n^{(m)}(X_m^n))_{m \in \mathcal{I}_{M}})))_{k \in \mathcal{I}_{K}} \\
\le &(D_k)_{k \in \mathcal{I}_{K}}.
\end{IEEEeqnarray*}
Furthermore, for any $\gamma > 0$ and sufficiently large $n$, we have
\begin{IEEEeqnarray*}{rCl}
\sum_{m \in A} R_m &\ge &\sum_{m \in A} R(\phi_n^{(m)}) - |A|\gamma \\
&\gevar{(a)} &\sum_{m \in A} \plimsup \frac{1}{n} \ln \frac{1}{P_{Z_n^{(m)}}(Z_n^{(m)})} - 2|A|\gamma \\
&\gevar{(b)} &\sum_{m \in A} \plimsup \frac{1}{n} \ln \frac{P_{Z_n^{(m)}|X_m^n}(Z_n^{(m)}|X_m^n)}{P_{Z_n^{(m)}}(Z_n^{(m)})} - 2M\gamma \\
&= &\sum_{m \in A} \overline{I}(\bm{X}_m; \bm{Z}^{(m)}) - 2M\gamma \\
&\gevar{(c)} &\sum_{m \in A} \overline{I}(\bm{X}_m; \bm{Z}^{(m)}) - \underline{I}^{|A|}((\bm{Z}^{(m)})_{m \in A}) - \underline{I}((\bm{Z}^{(m)})_{m \in A}; \bm{S}, (\bm{Z}^{(m)})_{m \in \mathcal{I}_{M} \backslash A}) - 2M\gamma
\end{IEEEeqnarray*}
where (a) follows from \cite[Lemma 2.6.2]{MSC:Han200300}, (b) from the fact that $P_{Z_n^{(m)}|X_m^n}(Z_n^{(m)}|X_m^n) \le 1$ and $|A| \le M$, and (c) from the nonnegativity of $\underline{I}^{|A|}((\bm{Z}^{(m)})_{m \in A})$ and $\underline{I}((\bm{Z}^{(m)})_{m \in A}; \bm{S}, (\bm{Z}^{(m)})_{m \in \mathcal{I}_{M} \backslash A})$. Since $\gamma$ is arbitrary, we have
$$
\sum_{m \in A} R_m \ge \sum_{m \in A} \overline{I}(\bm{X}_m; \bm{Z}^{(m)}) - \underline{I}^{|A|}((\bm{Z}^{(m)})_{m \in A}) - \underline{I}((\bm{Z}^{(m)})_{m \in A}; \bm{S}, (\bm{Z}^{(m)})_{m \in \mathcal{I}_{M} \backslash A}).
$$
Therefore, the converse part is established.
\end{proofof}

\bibliographystyle{IEEEtran} 
\bibliography{IEEEabrv,ISMSC}

\end{document}